\documentclass[aps,twocolumn,preprintnumbers,floatfix,showpacs,prx]{revtex4-1}

\usepackage{amsmath,amssymb}
\usepackage{dcolumn}
\usepackage{braket}
\usepackage{dsfont}
\usepackage{hhline}
\usepackage{graphicx}
\usepackage{xfrac}
\usepackage{paralist}

\usepackage{color}
\definecolor{darkblue}{rgb}{0.,0.,0.4}
\definecolor{darkred}{rgb}{0.5,0.,0.}
\usepackage[pdftex,colorlinks=true,linkcolor=darkblue,citecolor=darkred,urlcolor=blue]{hyperref}

\begin{document}

\title{A New Kind of Topological Quantum Order:\\ A Dimensional Hierarchy of Quasiparticles Built from Stationary Excitations}
%Stationary Excitations and their Mobile Composites}

\author{Sagar Vijay}
\author{Jeongwan Haah}
\author{Liang Fu}
\affiliation{Department of Physics, Massachusetts Institute of Technology, Cambridge, MA 02139}

\begin{abstract}
We introduce exactly solvable models of interacting (Majorana) fermions in $d \ge 3$ spatial dimensions that realize a new kind of topological quantum order, building on a model presented in ref. \cite{vijay}. 
These models have \emph{extensive} topological ground-state degeneracy
and a hierarchy of point-like, topological excitations that are only free to move within sub-manifolds of the lattice.
In particular, one of our models has fundamental excitations that are completely stationary. 
To demonstrate these results, we introduce a powerful polynomial representation of commuting Majorana Hamiltonians.
Remarkably, the physical properties of the topologically-ordered state are encoded in an algebraic variety,
defined by the common zeros of a set of polynomials over a finite field.
This provides a ``geometric" framework for the emergence of topological order.
\end{abstract}

\maketitle
Topological phases of matter are remarkable quantum states with quantized properties 
that are stable under local perturbations and can only be measured by non-local observables  \cite{wenbook}.     
The most celebrated example is the fractional quantum Hall state, discovered more than 
thirty years ago \cite{tsui}. The field of topological matter has now become an exciting research frontier  
at the crossroads between theoretical physics, quantum information and material science. 

Our theoretical understanding of topological matter is largely built on topological quantum field theory (TQFT) \cite{witten}. 
In this framework, the action of quantum fields in a  space-time manifold is independent of its metric, but 
depends crucially on its topology. Canonical quantization of these fields in a multiply-connected space yields a {\it finite-dimensional} 
Hilbert space, describing the degenerate ground-states of topological matter. %which describes the ground state degeneracy of topological matter.
Wilson lines describe the world-lines of 
quasi-particle excitations, and the expectation value of 
``knotted'' Wilson lines determines the quasi-particle braiding statistics.  
A hallmark of topologically-ordered states in two dimensions is the presence of {\it mobile} quasi-particles with fractional statistics, or anyons \cite{anyon}. 

Exactly solvable models often provide ideal playgrounds and valuable insights in theoretical studies of topological phases. 
In the past, a wide array of non-chiral topological phases in two dimensions 
have been obtained in spin models \cite{kitaev, levinwen}, whose universal properties are captured by topological quantum field theories.   
Recently, an exotic quantum phase with { extensive} topological ground state degeneracy was discovered by Haah in three-dimensional (3D) spin models \cite{Haah_Code}.  
A remarkable property of %Haah's model
this phase is that \emph{all} topological excitations are {\em strictly localized} in space,
a feature which lies beyond the paradigm of topological quantum field theory.

In this work, we introduce a wide range of translationally-invariant, solvable Hamiltonians of interacting Majorana fermions
that exhibit a new kind topological quantum order. These models have extensive topological degeneracy and a hierarchy of topological excitations that are only free to move within sub-manifolds of the full lattice.  In one particular Hamiltonian in $d = 3$ spatial dimensions, the fundamental excitations are strictly localized, while {composites} of these excitations are free to move along one- and two-dimensional surfaces.  The fundamental excitations are termed ``fractons",
as they behave as fractions of a mobile particle. 

To systematically search for these models, compute their ground-state degeneracy on a $d$-dimensional torus and study their excitations,
we introduce a purely algebraic description of commuting Majorana Hamiltonians. 
We demonstrate that on a $d$-dimensional lattice with a two-site basis and a single interaction term per unit cell,
an ideal Majorana Hamiltonian \emph{generally} exhibits extensive topological degeneracy.
%{\bf JH: under the assumption that there is a single interaction term per unit cell}
%Remarkably, while the fundamental excitations in these models are strictly localized, \emph{composites} of these excitations are free to move within sub-manifolds of the full lattice.  We term these frozen fundamental excitations that behave as fractions of a mobile particle, ``fractons".
We emphasize that each of our models may be written in terms of complex fermions by choosing appropriate pairings of Majorana fermions over the entire lattice.  Our models also admit a \emph{local} mapping to a boson model with identical topological degeneracy and a similar dimensional hierarchy of excitations, after projecting out half of the Hilbert space.  We note that one of our models has similar phenomenology to a spin model studied in ref. \cite{Bravyi, Chamon}.

Our approach to studying ideal Majorana Hamiltonians
provides a novel geometric framework for topological order, 
beyond topological quantum field theory.
Remarkably, a commuting Majorana Hamiltonian on a torus specifies an \emph{algebraic variety}
-- defined as the common zeros of a collection of polynomials over a finite field -- 
that encodes all physical properties of the topologically-ordered state.
While a TQFT assigns a ground-state sector to an isotopy class of smooth,
closed curves on a manifold, our models associate ground-state sectors with curves based on finer equivalence relations,
resulting in extensive topological degeneracy in dimensions $d \ge 3$.
We emphasize that our models are distinct from the exotic phase realized by Haah's code \cite{Haah_Code},
due to the presence of mobile composite %{\bf JH: composite}
topological excitations.

\section{Overview}
\label{sec:overview}

Due to the length of this paper, we begin with a detailed summary of our findings.
We consider exactly solvable Hamiltonians of interacting Majorana fermions that realize exotic forms of topological order.
On a $d$-dimensional lattice with a basis, these Hamiltonians will be the sum of a single type of local operator over all lattice sites
\begin{align}\label{eq:Hamiltonian}
H = -\sum_{m}\mathcal{O}_{m}
\end{align}
so that all operators mutually commute and square to the identity, i.e., 
\begin{align}
[\mathcal{O}_{m},\mathcal{O}_{n}] = 0 , \\ (\mathcal{O}_{n})^{2} = +1.
\end{align}
The operator $\mathcal{O}_{n}$ is required to be a product of an even number of Majorana fermions,
so that the fermion parity of the entire system is conserved.
A ground state $\ket{\Psi}$ of (\ref{eq:Hamiltonian}) will satisfy the constraint that 
\begin{align}
\mathcal{O}_{m}\ket{\Psi} = \ket{\Psi}, \label{gs}
\end{align}
for all $m$.

%We use a purely algebraic description of the ideal Majorana Hamiltonian in order to study a large range of Majorana models with topologically-ordered ground-states.  
In Section~\ref{sec:algebraic},
we introduce a purely algebraic approach to systematically search for topological order in commuting Majorana Hamiltonians (\ref{eq:Hamiltonian}).  
A similar approach has been used previously to study topological order in commuting Pauli Hamiltonians \cite{Haah_Commuting_Pauli}.
We represent the operator $\mathcal{O}$ appearing in (\ref{eq:Hamiltonian})
as a set of Laurent polynomials over the field $\mathbb{F}_{2}$,
which consists of two elements $\{0, 1\}$ with $\mathbb{Z}_{2}$ addition and multiplication.
%The degeneracy on a $d$-dimensional torus is calculated as the dimension of a quotient ring. 
We derive a mathematical condition for a set of such polynomials to represent a commuting Majorana Hamiltonian with topological order.
%We identify transformations on these polynomials that correspond to unitary and stable equivalence operations for the corresponding Hamiltonian. 
This polynomial representation enables us to  analytically determine the topological ground state degeneracy on a $d$-dimensional torus %as the dimension of a quotient ring  
and deduce properties of topological excitations using algebraic methods.

\begin{figure}
$\begin{array}{c}
\includegraphics[trim = 322 241 322 160, clip = true, width=0.31\textwidth, angle = 0.]{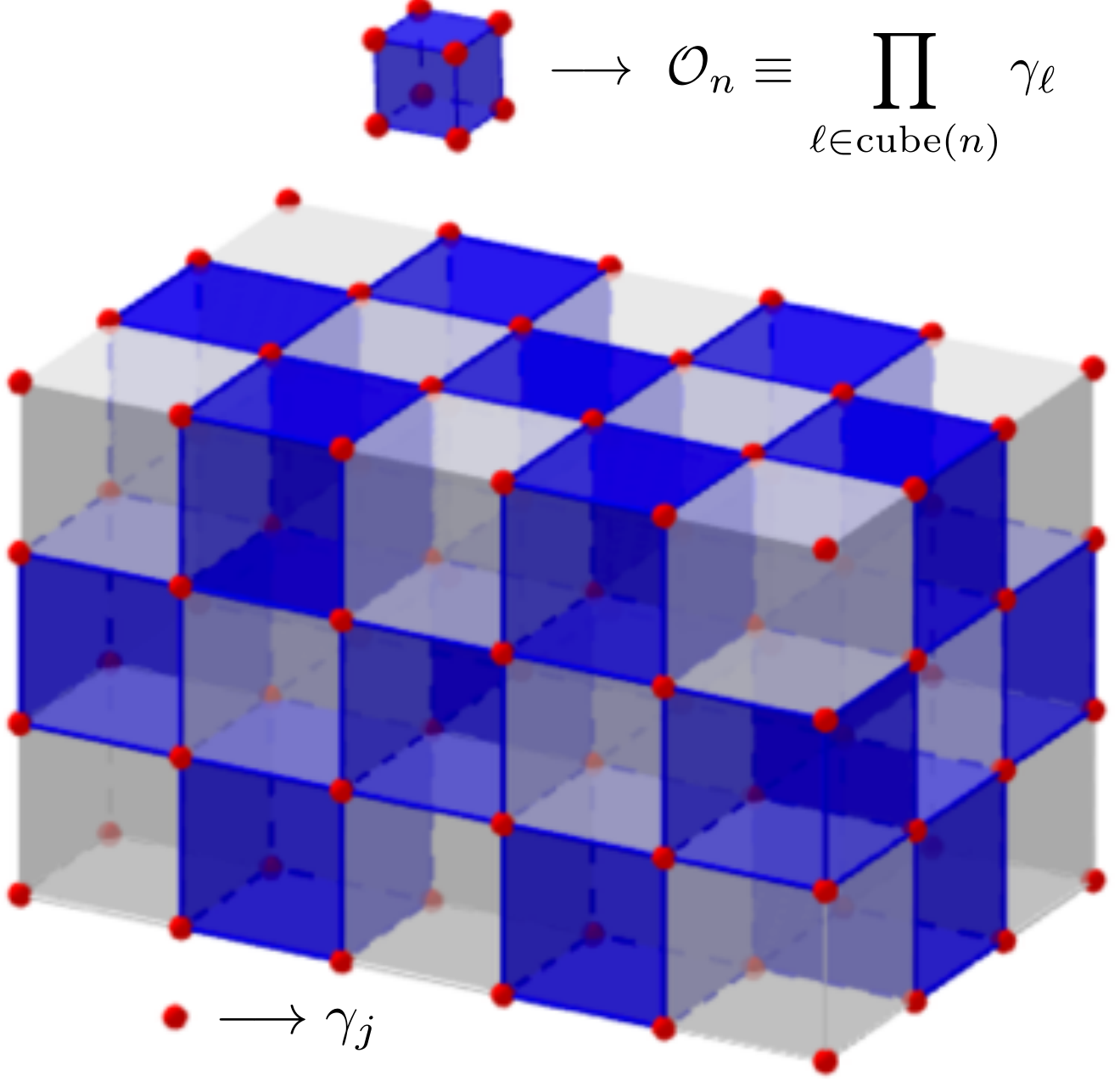}\\
\text{(a)}\\\\
\includegraphics[trim = 360 326 421 200, clip = true, width=0.18\textwidth, angle = 0.]{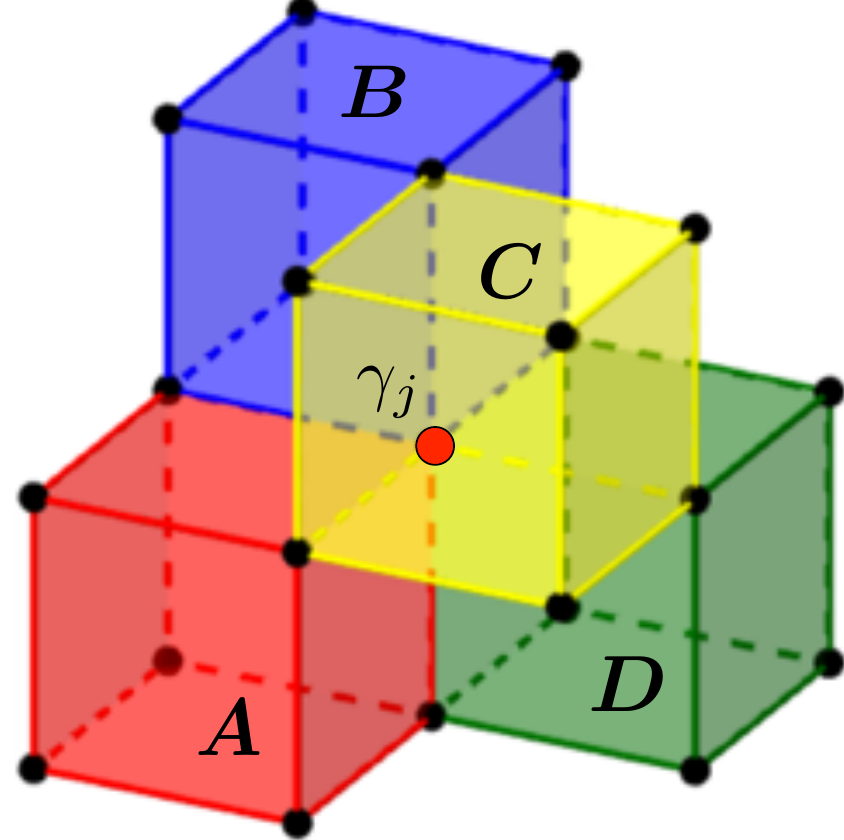}\\
\text{(b)}
\end{array}$
\caption{{\bf Majorana Cubic Model:}
The Majorana cubic model is defined on a cubic lattice, as in (a), with a single Majorana fermion per lattice site (colored red).
The operator $\mathcal{O}_{n}$ is the product of the $8$ Majorana fermions at the vertices of a cube.
The Hamiltonian is a sum of these local operators over every \emph{other} cube (colored blue) in a checkerboard pattern. 
As any pair of operators either share exactly one edge or none, all operators mutually commute. 
We choose to label the cubic operators $A$, $B$, $C$, and $D$ as shown in (b).
Acting with a single Majorana operator $\gamma_{j}$ creates these four excitations.}
\label{fig:Majorana_Cubic_Model}
\end{figure}

Using this polynomial approach, we demonstrate the following remarkable results.
First, a topologically-ordered commuting Majorana Hamiltonian on a lattice with a two-site basis
may be entirely specified by a single polynomial over $\mathbb{F}_{2}$.
The ground state degeneracy for such a Hamiltonian on a $d$-dimensional torus of size $L$, which we denote by $D_{0}$,
will \emph{generally} take the asymptotic form:
\begin{align}
\log_{2}D_{0} = c\,L^{d - 2} + O(L^{d-3})
\label{eq:Degeneracy_Scaling}
\end{align}
for some constant $c$.
We perform an exhaustive analysis and discover a class of commuting Majorana fermion models on a three-dimensional lattice with a two-site basis,
which exhibit {\it extensive} topological degeneracy of the form (\ref{eq:Degeneracy_Scaling}) with $d = 3$.

%A representation of the operator $\mathcal{O}$ defining each of these models is shown in Figure \ref{fig:3D_Models},
%along with the maximum ground-state degeneracy on an $L\times L\times L$ three-torus, for the indicated values of $L$.
%The first five models are shown to admit mobile topological excitations. 

Remarkably, despite being translationally invariant, our models admit fundamental point-like excitations that are strictly localized in space,
and cannot move without paying a finite energy cost to create additional excitations.
%Composites of these fundamental excitations, however, may be created with a local operator,
Composites of these fundamental excitations, however, %may be created with a local operator.  These composite, topological excitations behave as particles
are topological excitations that are free to move within {\it sub-manifolds} of the $d$-dimensional lattice. 
%{\bf JH: emphasis on mobile topological excitation?}
We term these fundamental excitations that behave as fractions of mobile particles, ``fractons."
Furthermore, we refer to bound states of fractons that can only move freely along an $n$-dimensional manifold as ``dimension-$n$" particles.
In particular, a dimension-2 particle can be an anyon with well-defined fractional statistics.  

To motivate further study of ideal Majorana Hamiltonians,
we now describe in detail the phenomenology of fracton excitations and their composites in the simplest of our models, the Majorana cubic model.
As shown in Figure~\ref{fig:Majorana_Cubic_Model}(a),
here the operator $\mathcal{O}_{n}$ is the product of the eight Majorana fermions at the vertices of a cube.
The Hamiltonian is simply the sum of these operators over a face-centered-cubic (fcc) array of cubes, forming a three-dimensional checkerboard.
Since adjacent cubes share a common edge with two vertices, operators  $\mathcal{O}_{n}$ on different cubes are mutually commuting,
and their common eigenstate defines the ground state.
For convenience in later analysis, we choose to identify four species of cube operators -- $A$, $B$, $C$, and $D$ -- 
as shown in Figure~\ref{fig:Majorana_Cubic_Model}(b).

 \begin{table}
\begin{tabular}{|c|c|c|c|}
  \hline
  $\begin{array}{c}\\\mathbf{\underline{Excitation}}\\ \\\end{array}$ & \underline{\bf Type} & \underline{\bf Statistics} & \underline{\bf Operator}\\
    \hhline{|=|=|=|=|}
    $\begin{array}{c} \\ ABCD \\ \\\end{array}$ & Majorana & Fermion & $\gamma$\\ 
      \hline
      $\begin{array}{c} \\ AA, BB, \\CC, DD\\ \\\end{array}$ & Dim.-2 Anyon & Boson & $\begin{array}{c}\text{Pair of Adjacent}\\ \text{Wilson Lines}\end{array}$ \\
      \hline
      $\begin{array}{c} \\ AB,\, AC,\\ AD,\, 
      BC,\\ BD,\, CD \\ \\\end{array}$ & Dim.-1 Particle & -------- & $\begin{array}{c}\text{Single Wilson}\\ \text{Line}\end{array}$ \\
      \hline
      $\begin{array}{c}\\ A,\,B,\,C,\,D \\ \\ \end{array}$ & Fracton & -------- & Membrane\\
      \hline
 \end{tabular}
 \caption{Hierarchy of excitations in the Majorana cubic model. 
 The fundamental cube excitation is a fracton,
 while two-fracton bound-states can behave as particles that are either free to move along one- or two-dimensional surfaces.
 The operator that creates each type of excitation is indicated.}
 \label{tab:Hierarchy}
 \end{table}
 
A fundamental excitation in the Majorana cubic model is obtained when the eigenvalue of a cube operator $\mathcal{O}_n$ is flipped. 
The product of $\mathcal{O}_n$ over all cubes of a single type ($A$, $B$, $C$, or $D$)
is equal to the fermion parity $\Gamma$ of the entire system and is fixed.
\begin{align}\label{Constraints}
\Gamma 
= \prod_{p \in A}{\mathcal{O}}_{p} 
= \prod_{p \in B}{\mathcal{O}}_{p} 
= \prod_{p \in C}{\mathcal{O}}_{p} 
= \prod_{p \in D}{\mathcal{O}}_{p}, 
\end{align}
Therefore, a single cube-flip excitation cannot be created alone, and is a topological excitation. 
Remarkably, the fundamental cube excitation in this model is completely \emph{immobile},
as we observe through the following physical argument.
In the cubic model, acting on the ground-state with a single Majorana fermion
flips the eigenvalues of four adjacent cube operators, as shown in Figure~\ref{fig:Majorana_Cubic_Model}(b).
This four-cube excitation may trivially move by acting with a Majorana bilinear.
If the fundamental cube excitation were mobile,
then it would be possible to move it in any arbitrary direction, as the cube operator itself preserves all lattice symmetries.
In this case, the cube excitation would have well-defined (fermion or boson) statistics,
and a four-cube bound-state could never be a \emph{fermion}.
Therefore, it must be the case that the fundamental cube excitation is frozen.
A rigorous proof of the immobility of the fundamental excitation is given in Section~\ref{sec:fracton}
using the polynomial representation of the ideal Majorana Hamiltonian.

%Roughly speaking, this model behaves as copies of $Z_{2}$ gauge theory coupled to additional statistical gauge fields.

\begin{figure}[t]
$\begin{array}{c}
\,\,\,\,\,\includegraphics[trim = 329 250 259 250, clip = true, width=0.29\textwidth, angle = 0.]{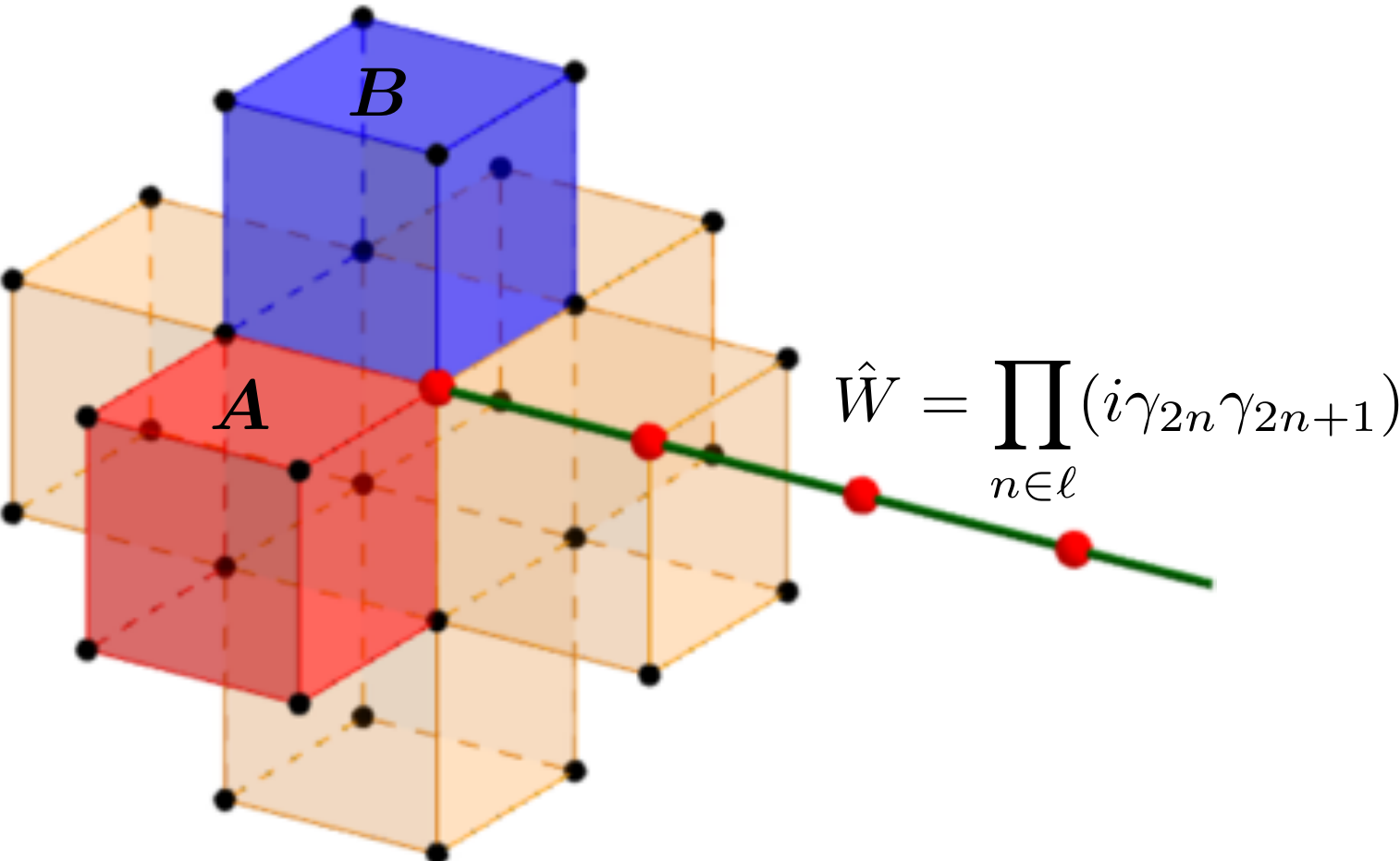}\\
\text{(a)}\\
\includegraphics[trim = 650 421 870 310, clip = true, width=0.29\textwidth, angle = 0.]{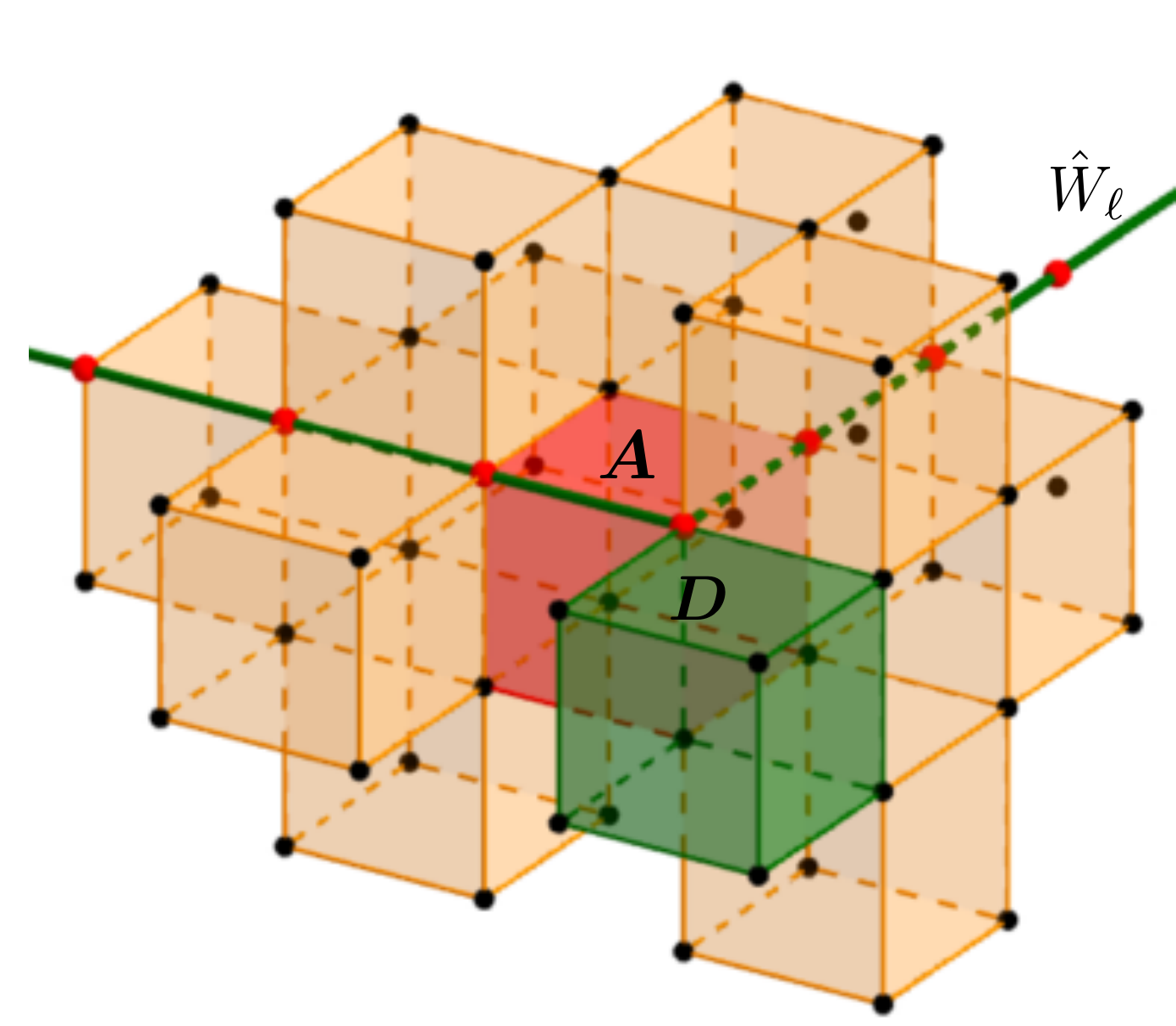}\\
\text{(b)}
\end{array}$
%{\bf UNFINISHED FIGURE}
\caption{{\bf Dimension-1 Particle}:
Excitations (colored) may be created by acting with Wilson line operators.
In (a), a straight Wilson line creates pairs of dimension-1 particles at the endpoints. The dimension-1 particle may hop freely in the direction of the Wilson line, by acting with Majorana bilinear terms.  Remarkably, the dimension-1 particle cannot hop in any other direction without creating additional excitations.  Introducing a ``corner" in the Wilson line, as in (b), creates an additional topological excitation localized at the corner.  }
\label{fig:Linear_Anyon}
\end{figure}

We now analyze the fracton bound-states in the Majorana cubic model in detail,
along with the mutual statistics of the excitations.
Using the labeling of the cube operators shown in Figure~\ref{fig:Majorana_Cubic_Model}(b),
we find the hierarchy of quasiparticles shown in Table~\ref{tab:Hierarchy} in the Majorana cubic model.
The fundamental fracton excitation appears at the \emph{corners} of membrane-like operators
and may only be created in groups of four.
Two-fracton bound-states can form dimension-1 particles or dimension-2 anyons.
Remarkably, a dimension-2 anyon has $\pi$ mutual statistics with a fracton lying in its plane of motion.
As a result, while the fracton is immobile, its presence may be detected by a braiding experiment.
Furthermore, the exact location of a single fracton within a finite volume $V$ may be determined by braiding dimension-2 anyons in the three mutually orthogonal planes around the boundary $\partial V$. In this way, the exact quasiparticle content within $V$ is effectively encoded ``holographically", and may be determined by $\sim O(\ell)$ braiding experiments, where $\ell$ is the linear size of a box bounding $V$.

We now proceed to explore the hierarchy of excitations in detail.

{\bf Dimension-1 Particle:}
The dimension-1 particle may be created by acting with a single  Wilson line operator,
defined by the product of the Majorana operators along a straight path $\ell$.
Up to an overall pre-factor of $\pm 1,\pm i$, we write the Wilson line operator as
\begin{align}
\hat{W}_{\ell} \propto \prod_{n \in \ell}\gamma_{n}.
\end{align} 
As shown in Figure~\ref{fig:Linear_Anyon}(a),
the straight Wilson line anti-commutes with two cube operators at each of its endpoints;
the two cube excitations at a given endpoint are of different types.
As a result, $\hat{W}_{\ell}$ creates pairs of excitations of the form $AB$, $AC$, $AD$, $BC$, $BD$, or $CD$.
Remarkably, these two-fracton bound-states are only free to move along a line,
by simply extending the Wilson line operator $\hat{W}_{\ell}$ by acting with a Majorana bilinear along the path $\ell$.
If we try to move this two-fracton bound-state in a plane,
we must introduce a corner in the Wilson line,
which localizes an additional \emph{topological} excitation at the corner, as shown in Figure~\ref{fig:Linear_Anyon}(b); the excitation cannot be removed by the action of any local operator.
As the pattern of excitations produced by a Wilson line $\hat{W}_{\ell}$ is sensitive to the \emph{geometry} of $\ell$,
the two-fracton bound-states $AB$, $AC$, $AD$, $BC$, $BD$, and $CD$ are restricted to move along a line and behave as dimension-1 particles.
%{\bf JH: Perhaps, it is worthnoting here that any two-fracton bound state is a topological excitation. If not, the fracton would be mobile.
%Since the kink creates a topological excitation it cannot be removed however one decorates the corner.}
We emphasize that they cannot move
in a higher-dimensional space without creating additional cube excitations.

\begin{figure}[t]
$\begin{array}{c}
\includegraphics[trim = 300 210 259 220, clip = true, width=0.29\textwidth, angle = 0.]{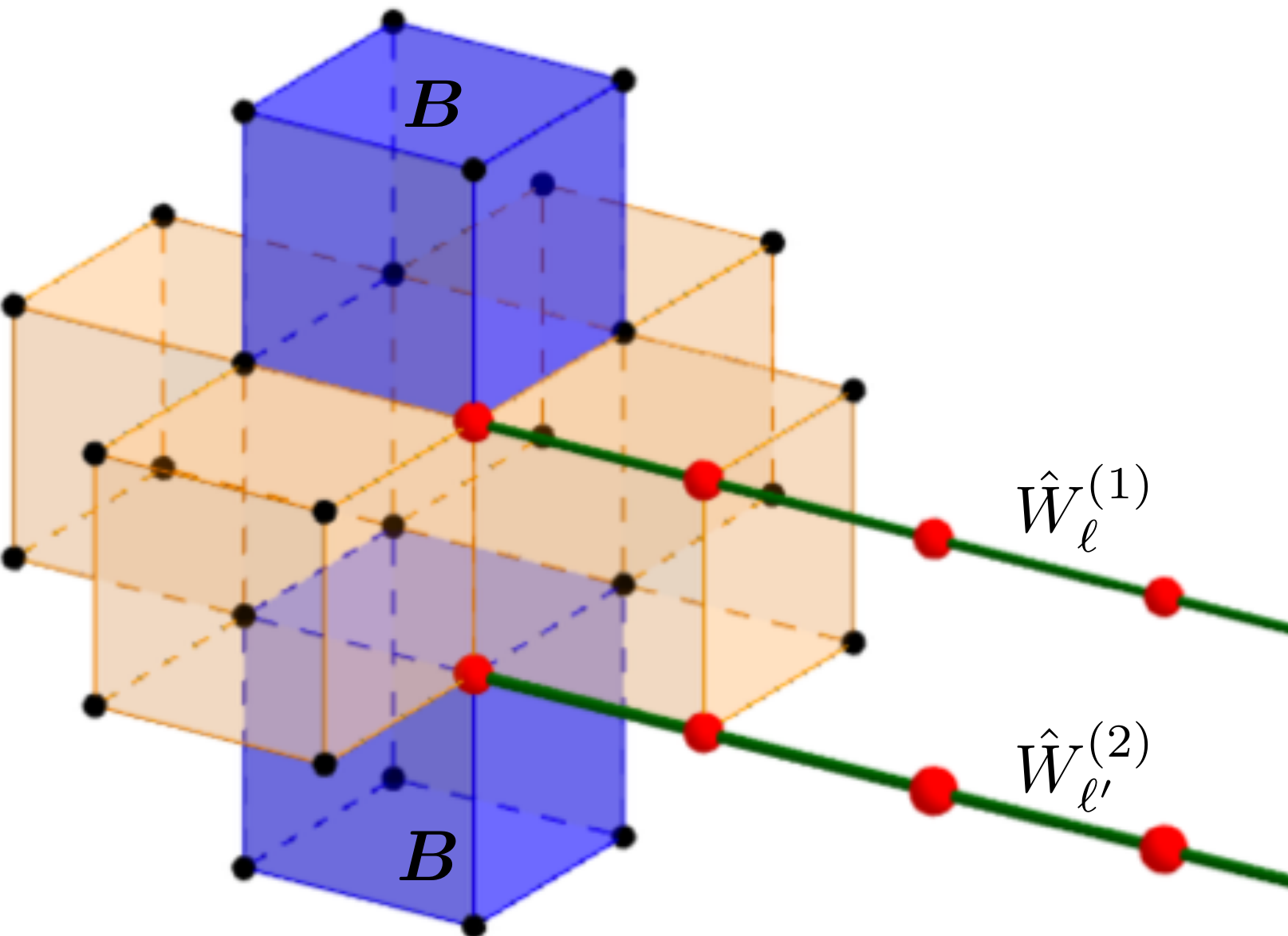}\\
\text{(a)}\\\\
\includegraphics[trim = 713 482 800 310, clip = true, width=0.29\textwidth, angle = 0.]{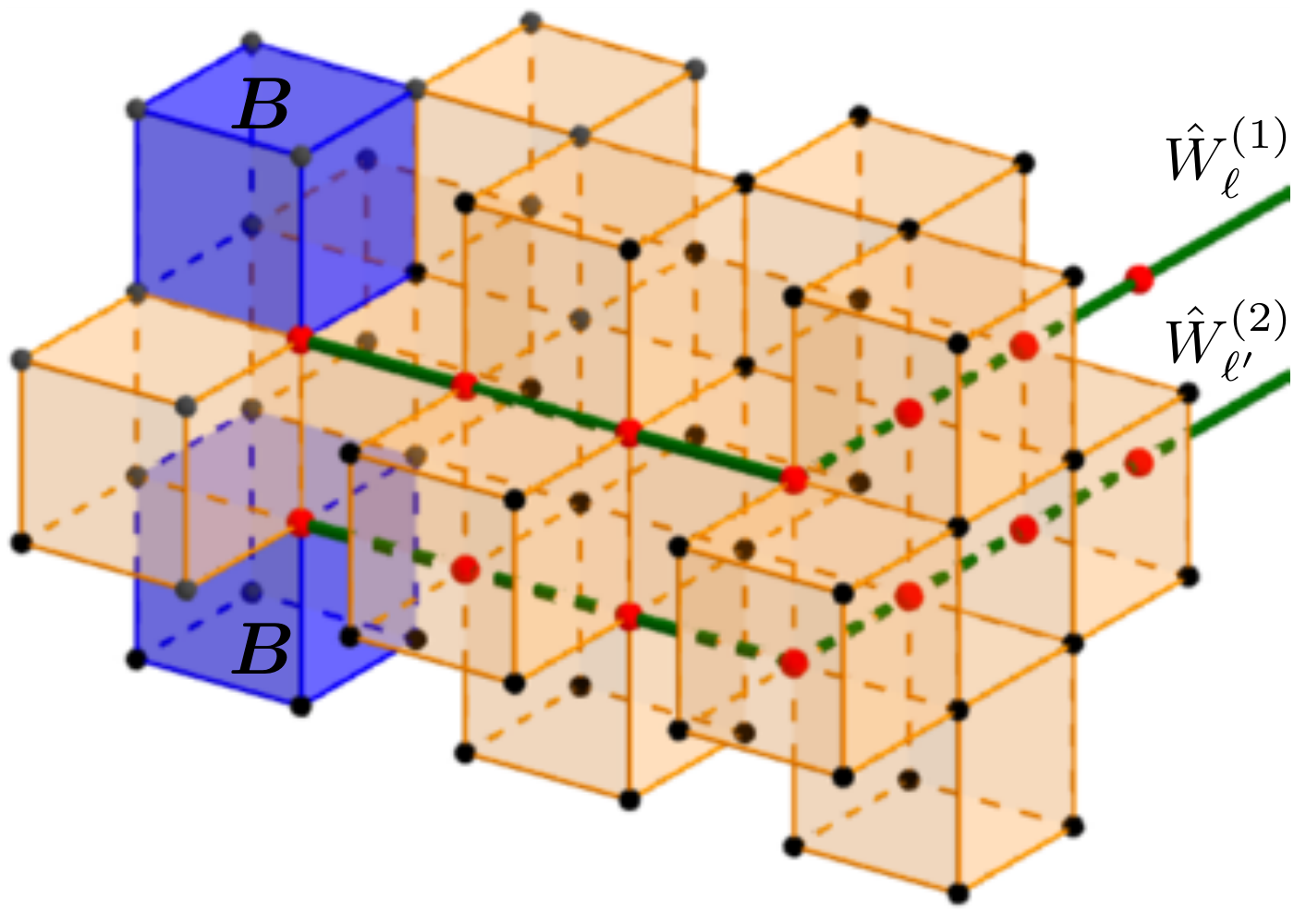}\\
\text{(b)}\\\\
\includegraphics[trim = 210 250 400 150, clip = true, width=0.29\textwidth, angle = 0.]{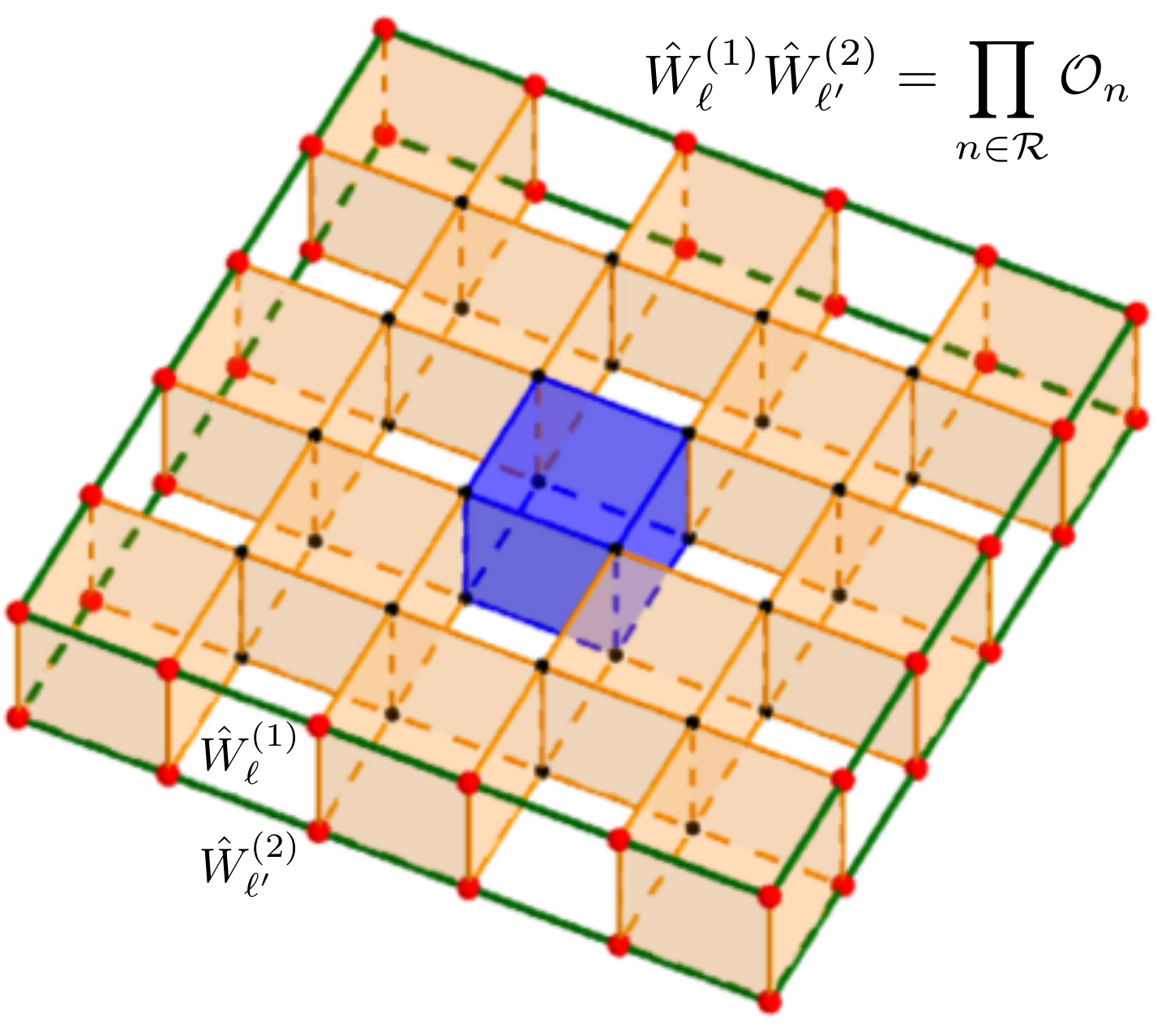}\\
\text{(c)}
\end{array}$
\caption{{\bf Dimension-2 Anyon}:
Acting with two adjacent Wilson line operators  $\hat W _1$ and $\hat W_2$
creates pairs of excitations at the endpoints of the same type ($AA$, $BB$, $CC$ or $DD$).
These two-fracton excitations are free to move in a two-dimensional plane
orthogonal to the shortest line segment connecting the pair of Wilson lines. %{\bf JH: $\leftarrow$ tweaked}
Furthermore, in (b) we may detect a fracton (colored blue) by braiding a dimension-2 anyon around a closed loop enclosing the fracton.
As the braiding operator, a pair of closed Wilson line operators $\hat{W}_{1}\hat{W}_{2}$,
is equal to the product of the enclosed cube operators as shown above. Therefore,
the braiding produces an overall minus sign if an odd number of fractons are enclosed.}
\label{fig:Planar_Anyon}
\end{figure}

\begin{figure}[t]
\includegraphics[trim = 240 227 320 140, clip = true, width=0.32\textwidth, angle = 0.]{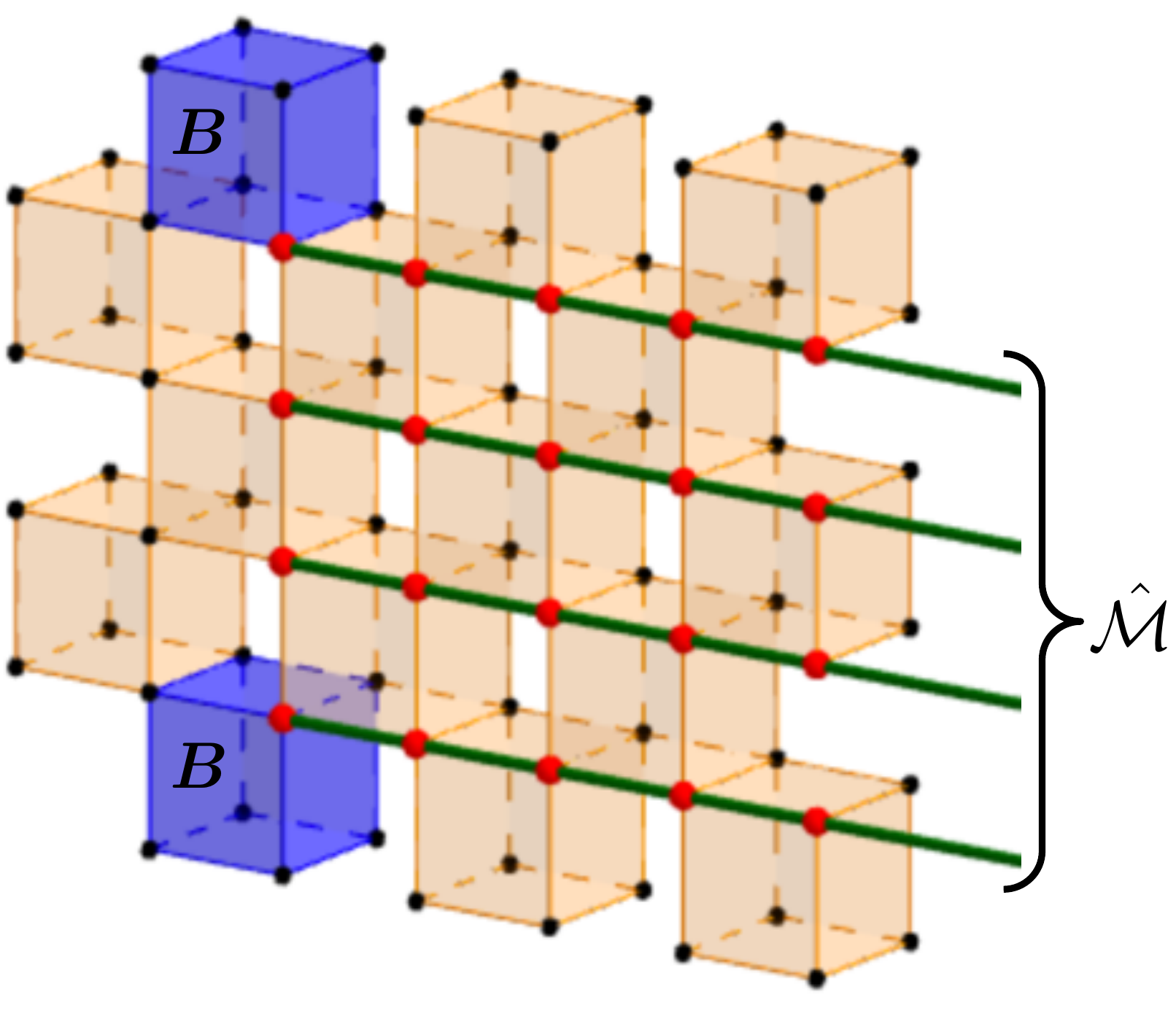}
\caption{{\bf Membrane Operator \& Fracton Excitations}:
Acting with a product of Majorana operators on a surface $\Sigma$
creates localized excitations at the corners of the boundary $\partial\Sigma$ as shown above. }
\label{fig:Membrane}
\end{figure}

{\bf Dimension-2 Anyon:}
Acting with a pair of adjacent Wilson lines $\hat{W}^{(1)}_{\ell}$ and $\hat{W}^{(2)}_{\ell'}$
along parallel paths $\ell$ and $\ell'$, respectively,
also creates a pair of two-fracton bound-states localized at the ends,
as shown in Figure~\ref{fig:Planar_Anyon}(a).
At each end of the path, however, the operator $\hat{W}^{(1)}_{\ell}\hat{W}^{(2)}_{\ell'}$
now creates pairs of cube excitations of the \emph{same} type ($AA$, $BB$, $CC$ or $DD$).
These two-fracton bound-states, where each fracton is of the same type,
are allowed to move freely in the two-dimensional plane orthogonal to the shortest line segment connecting the two paths $\ell$ and $\ell'$
without creating additional excitations; this is shown in Figure \ref{fig:Planar_Anyon}(b).
We note that detailed geometric features of a \emph{single} Wilson line, such as the presence of sharp corners, 
determine the pattern of excitations created from the ground-state.
However, when acting with an appropriate \emph{pair} of adjacent Wilson lines,
the excitations created at the sharp corners may be annihilated.
Therefore, a pair of adjacent Wilson lines may be deformed in the plane with no energy cost. 
We conclude that the $AA$, $BB$, $CC$ and $DD$ two-fracton bound-states are dimension-2 anyons.

Braiding a dimension-2 anyon around a closed loop in the plane
is equivalent to acting with the product of cube operators within the two-dimensional region enclosed by the loop;
this is shown for a particular choice of loop in Figure~\ref{fig:Planar_Anyon}(c).
As a result, braiding a dimension-2 anyon around a closed loop enclosing a single fracton in the plane
produces an overall minus sign.
The ability to detect a fracton with a dimension-2 anyon 
produces non-trivial mutual statistics between the dimension-2 anyon and other particles
in the excitation spectrum of the Majorana cubic model.
First, a dimension-2 anyon has $\pi$ mutual statistics with any dimension-1 particle in the same plane,
as braiding the dimension-2 anyon in a closed loop
will only detect one of the two fractons that make up the dimension-1 particle.
Furthermore, the dimension-2 anyon has 
$\pi$ mutual statistics with dimension-2 anyons that are free to move in adjacent, parallel planes.

{\bf Fractons and Membrane Operators:}
Acting with Majorana operators on a flat, two-dimensional membrane $\Sigma$ 
creates fracton excitations at the \emph{corners} of the boundary of $\Sigma$,
as shown in Figure \ref{fig:Membrane}.
We write the membrane operator up to an overall pre-factor of $\pm 1, \pm i$ as
\begin{align}
\hat{\mathcal{M}} \propto \prod_{n \in \Sigma}\gamma_{n}.
\end{align}
For a rectangular membrane in the $x$-$y$ plane,
the boundary $\partial\Sigma$ is a closed, rectangular loop with dimensions $\ell_{x}$ and $\ell_{y}$.
We note that if $\ell_{x}$ and $\ell_{y}$ are both even,
then the fracton excitations created at the corners of $\partial\Sigma$ will all be of the same type.
Alternatively, if $\ell_{x}$ is odd and $\ell_{y}$ is even,
then the pairs of fracton excitations separated in the $y$-direction will be of the same type,
while fractons separated in the $x$-direction will be distinct.  

%\begin{figure}
%\includegraphics[trim = 200 450 600 100, clip = true, width=0.21\textwidth, angle = 0.]{Cube.pdf}
%\caption{The cubic operator acts on the eight Majorana fermions on the vertices of the cube.
%The lattice translation operators are shown above along with the two-site unit cell,
%colored in blue and red, respectively.}
%\label{fig:Cube_Unit_Cell}
%\end{figure}

%\begin{figure}
%\includegraphics[trim = 106 505 601 90, clip = true, width=0.31\textwidth, angle = 0.]{FCC_Model.pdf}
%\caption{We consider a single complex fermion per site on an fcc lattice. The Hamiltonian for }
%\label{fig:FCC_Model}
%\end{figure}

{\bf Extensive Topological Degeneracy:}
Using the algebraic representation of the Majorana cubic model,
we compute its ground-state degeneracy $D_{0}$ to be 
\begin{align}
\log_{2}D_{0} = 3L-3
\end{align}
on an $L\times L\times L$ three-torus, with periodic boundary conditions imposed in the $x$, $y$, and $z$ directions, with each cube having unit side length. 
%with the choice of translation vectors shown in Figure~\ref{fig:Cube_Unit_Cell}.
%{\bf JH: It is possible to compute the degeneracy using the usual counting,
%by inserting $(\bar x y z)^L-1, (x \bar y z)^L -1, (x y \bar z)^L-1$ in place of $x^L-1,y^L-1,z^L-1$.}
Pairs of string-like Wilson loop operators wrapping non-trivial cycles of the torus
-- corresponding to tunneling dimension-2 anyons -- 
distinguish the ground-state sectors.
As the number of distinct dimension-2 anyons grows linearly with system size,
the ground-state degeneracy is necessarily extensive.  

We emphasize that the algebraic approach 
allows us to systematically search for topologically-ordered, ideal Majorana Hamiltonians,
rigorously characterize the nature of excitations,
and calculate the ground-state degeneracy in a wide range of Majorana models using techniques in algebraic geometry.
As a result, the next two sections of this work introduce and focus on
the polynomial representation of ideal Majorana Hamiltonians 
and draw broad conclusions based on this representation.
In Section \ref{sec:extensive},
we present the 6 distinct three-dimensional Majorana models with %{\bf JH: Did we agree to use the term ``nearest''?}
nearest-neighbor interactions that are topologically-ordered.  In particular, one of our models, which may naturally be written in terms of complex fermions on an fcc lattice, has a fundamental excitation that may only freely move along a line in the (1,1,1) direction.  

We conclude, in Section~\ref{sec:fracton}, with a proof of the presence of fractons in the Majorana cubic model,
and briefly outline the phenomenology of excitations in the remaining models.  %We note that after completing this work, we have become aware of a spin model with similar phenomenology, studied in ref. \cite{Bravyi, Chamon}.
%{\bf JH: Maybe we can come back to this paragraph after everything is written in the corresponding section.}

\section{Topological Order in Commuting Majorana Hamiltonians}
\label{sec:algebraic}

In this section, we introduce a representation of the operators in the ideal Majorana Hamiltonian (\ref{eq:Hamiltonian})
as a vector of Laurent polynomials over the finite field $\mathbb{F}_{2}$.
The algebraic representation provides an important starting point for studying and classifying Majorana Hamiltonians.
We demonstrate that the following conditions, that
\begin{enumerate}[(i)]
\item All operators in the ideal Hamiltonian mutually commute, and
\item Degenerate ground-states of the Hamiltonian are locally indistinguishable
\end{enumerate}
may be phrased entirely in the polynomial representation.
The ground-state degeneracy of an ideal Majorana Hamiltonian (\ref{eq:Hamiltonian})
on the torus can be computed as the dimension of a quotient ring~\cite{Haah_Commuting_Pauli}.  

We demonstrate that an ideal Majorana Hamiltonian obeying (i) and (ii) on a lattice with a two-site basis and a single interaction term per unit cell
%{\bf JH: single interaction term}
may be specified by a \emph{single} polynomial over $\mathbb{F}_{2}$.
We use this result to systematically search for and characterize commuting Majorana Hamiltonians.
In $d = 3$ dimensions, we find 6 distinct, non-trivial models with
nearest-neighbor interactions,
extensive topological degeneracy, and a dimensional hierarchy of excitations.

\subsection{Algebraic Representation}
To study commuting Majorana Hamiltonians,
we represent the operator $\mathcal{O}$ appearing in Eq.~\eqref{eq:Hamiltonian}
as a polynomial over the field $\mathbb{F}_{2}$.
A similar mapping has been introduced in the context of Pauli Hamiltonians~\cite{Haah_Commuting_Pauli}.
Consider a $d$-dimensional lattice with translation operators
$\{\boldsymbol{t}_{1}, \ldots, \boldsymbol{t}_{d}\}$
and an $n$-site unit cell. We restrict $n$ to be an \emph{even} integer so that there is a well-defined number of complex fermions per lattice site. %{\bf JH: $2n$ because we want one complex fermion?}
We label the Majorana fermions within the unit cell at the origin as $\gamma_{j}$ for $j = 1, 2, \ldots, n$.
All other Majorana fermions on the lattice are obtained by acting with translation operators.

Any Hermitian operator acting on this lattice may be written as a sum of products of Majorana operators.
Formally, we may write a summand $\mathcal{O}$ as
\begin{align}
\mathcal{O} 
= \prod_{j = 1}^{n}  \prod_{\{n_{i}\}} 
  \left( \boldsymbol{t}_{1}^{n_{1}}\cdots\boldsymbol{t}_{d}^{n_{d}} \cdot \gamma_{j} \right)^{ c_{j} (n_1,\ldots,n_d) }
\label{eq:Operator}
\end{align}
with $n_{i} \in \mathbb{Z}$ and $c_{j}(n_{1},\ldots,n_{d}) \in \{0, 1\}$.
For simplicity, we have omitted the prefactor $\pm 1, \pm i$ in the expression for $\mathcal{O}$,
which plays no role in our analysis.
We introduce a purely algebraic representation of this operator
by noting that any product of translation operators may be written as a monomial,
e.g. $\boldsymbol{t}_1^{n_1}\cdots \boldsymbol{t}_d^{n_d}
\Longleftrightarrow
x_1^{n_1}\cdots x_d^{n_d}$.
In this way, the action of the translation group is naturally represented by monomial multiplication. 

Recall that distinct Majorana fermions anti-commute and that each Majorana operator squares to the identity.
Therefore at each site within a unit cell, the identity $\mathds{1}$ and $\gamma$ under multiplication form the group $\mathbb{Z}_{2}$,
with the two operators represented by the group elements $0$ and $1$, respectively. 
In this representation, the operator equality $\gamma^2 = \mathds{1}$
maps to the $\mathbb{Z}_{2}$ group addition $1+ 1=0$.  
This simple algebra of Majorana fermions allows us to write any product of Majorana operators as the {\em sum} of monomials
-- representing the location of each Majorana operator via the action of the translation group -- 
with $\mathbb{Z}_{2}$ coefficients.
As an example, consider a lattice with a single site per unit cell,
and the Majorana operator $\gamma$ at the origin.
A Majorana bilinear admits the following polynomial representation:
\begin{align}
\gamma \cdot 
  \left(\boldsymbol{t}_1^{m_1}\boldsymbol{t}_2^{m_2}\cdots\boldsymbol{t}_d^{m_d}\cdot \gamma \right)
 ~\Longleftrightarrow~
 1 + x_1^{m_1}x_2^{m_2}\cdots x_d^{m_d}.
\end{align}
In this notation, operator multiplication corresponds to polynomial addition with $\mathbb{Z}_2$ coefficients.

For the general case of a unit cell with $n$ sites,% {\bf JH: why don't we just use $n$ and say in the beginning $n$ is even.}
we represent a product of Majorana operators as a \emph{vector} of polynomials over $\mathbb{F}_{2}$,
with the $j$-th entry of the vector representing the action of the translation group on $\gamma_{j}$,
the $j$-th Majorana fermions in the unit cell at the origin.
For example, the operator \eqref{eq:Operator} may be written as
\begin{align}
S(x_{1}, . . .,x_{d}) = \sum_{\{ n_i \}}x_1^{n_1} \cdots x_d^{n_d}
\begin{pmatrix}
c_{1}(n_{1}, . . ., n_{d})\\
c_{2}(n_{1}, . . ., n_{d})\\
\vdots\\
c_{2n}(n_{1}, . . ., n_{d})
\end{pmatrix}
\label{eq:Operator_Polynomial}
\end{align}
Adopting the terminology in Ref.~\cite{Haah_Commuting_Pauli},
we refer to $S$ as the ``stabilizer map" for the remainder of this work.

\begin{figure}
$\begin{array}{cc}
\includegraphics[trim = 0 142 209 0, clip = true, width=0.2\textwidth, angle = 0.]{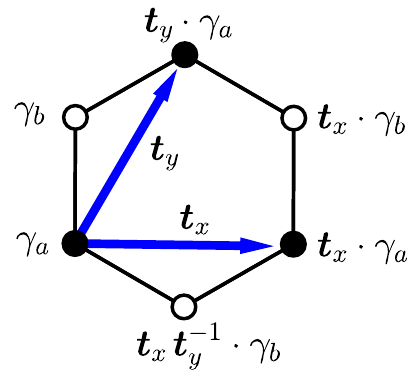} & \hspace{.1in}
\includegraphics[trim = 0 110 190 0, clip = true, width=0.2\textwidth, angle = 0.]{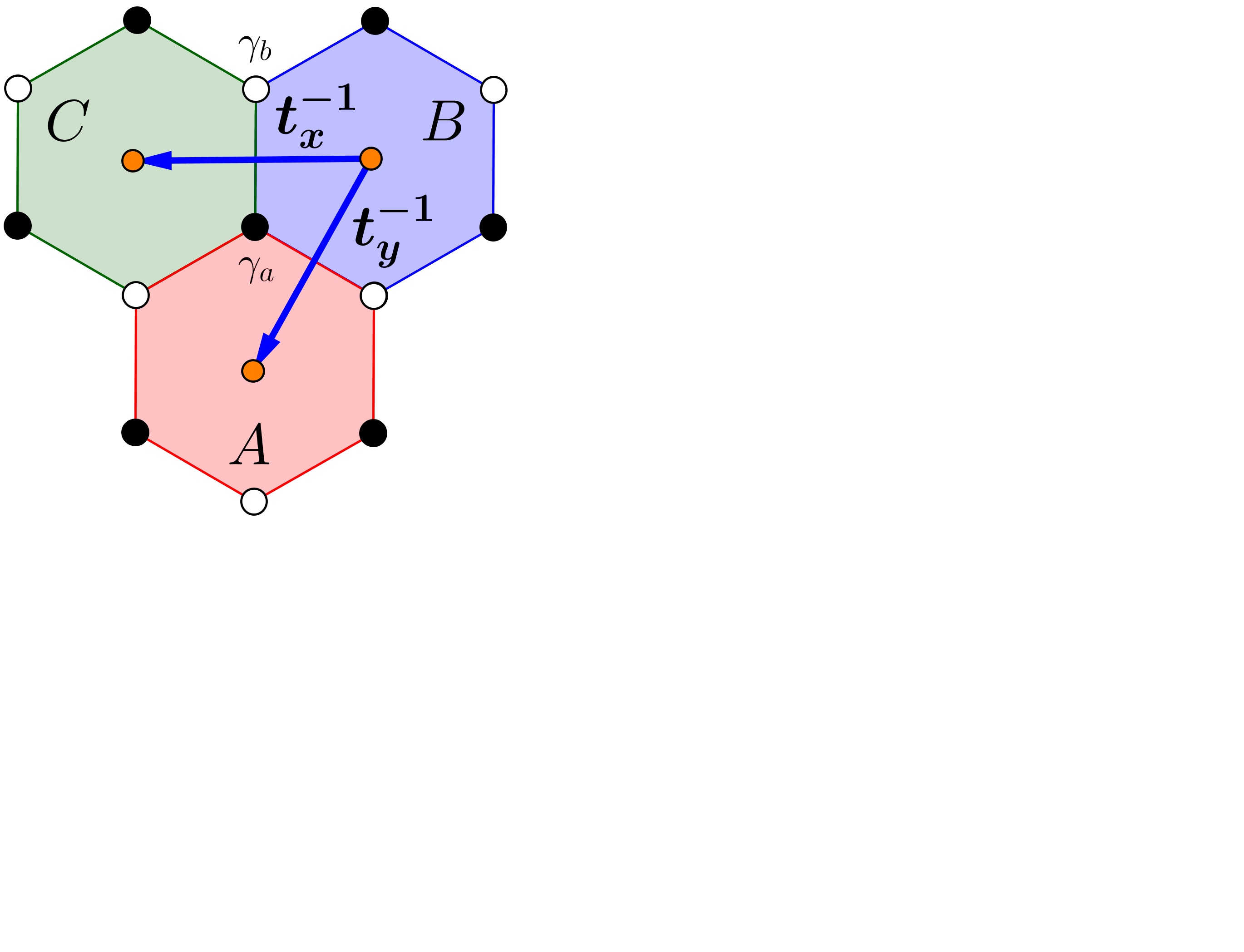}\\
\text{(a)} & \hspace{.1in} \text{(b)}
\end{array}$
\caption{The Majorana plaquette model, as studied in \cite{vijay}.
Consider a honeycomb lattice with a single Majorana fermion on each lattice site. We define an operator $\mathcal{O}_{p}$ as the product of the six Majorana fermions on the vertices of a hexagonal plaquette $p$, as shown in (a). 
The colored plaquettes in (b) correspond to the three distinct bosonic excitations ($A, B,$ or $C$)
that may each be created in pairs by acting with Wilson line operators.}
\label{fig:Plaquette_1}
\end{figure}

To illustrate the algebraic representation of operators in commuting Majorana Hamiltonians,
we present a concrete example.
Consider the Majorana plaquette model in Ref.~\cite{vijay},
which is defined on a two-dimensional honeycomb lattice with one Majorana fermion per site 
and a Hamiltonian of the form \eqref{eq:Hamiltonian} where $\mathcal{O}_p$
is the product of the six Majorana fermions at the vertex of a hexagonal plaquette $p$.
%{\bf JH: I see that $n$ is used to mean two different things.}
We show a single hexagonal plaquette on the lattice in Figure~\ref{fig:Plaquette_1}(a),
along with the Majorana fermions $\gamma_{a}$ and $\gamma_{b}$ within the two-site unit cell.
The corresponding stabilizer map $S(x,y)$ for the six-Majorana operator is given by:
\begin{align}
S(x,y)  = 
\begin{pmatrix}
1 + x + y\\
\\
1 + x + x\bar{y}
\end{pmatrix}.
\label{eq:smap-PlaquetteModel}
\end{align}
Here, we adopt the notation that $\bar{y} \equiv y^{-1}$, $\bar{x} \equiv x^{-1}$. 
As shown in Ref.~\cite{vijay},
this Hamiltonian exhibits a novel form of $\mathbb Z_2$ topological order
with fermion parity-graded excitations and exact anyon permutation symmetries.  

Next, we consider the action of an arbitrary operator $W$
on the ground state $\ket{\Psi}$ of the commuting Majorana Hamiltonian.    
When $W$ anticommutes with an operator $\mathcal{O}_{n}$ in the Hamiltonian,
it flips its eigenvalue and thus creates an excitation.  
We use a polynomial to record the locations of all excitations in the state $W\ket{\Psi}$;
each location is labeled by the translation vector connecting it to the origin.
 %As our Hamiltonian is translationally invariant, we may use the same translation vectors to label
 %the locations of the commuting operators $\{\mathcal{O}_{n}\}$ whose eigenvalues have been flipped through the action of $W$. 
%So far, we have used translation vectors to label lattice sites.
%{\bf JH: The notation of $S(\vec x)$ drives me to think of $S$ as a map from $\vec x$ to something....}
Specifically, for a Hamiltonian with stabilizer map $S(x_{1},\cdots,x_{d})$
and an arbitrary operator $W$ with a polynomial representation $P(W)$ of the form \eqref{eq:Operator_Polynomial},
we define the ``excitation map'' $E(x_{1},\ldots,x_{d})$ so that $E(x_{1},\ldots,x_{d}) \cdot P(W) \in \mathbb{F}_2[x_1^{\pm1}, \cdots, x_d^{\pm1}]$
describes the excitations created by $W$.
In the Supplemental Material~\cite{Supp_Mat},
we demonstrate that $E$ is simply given from the stabilizer map as follows: 
\begin{align}
E(x_{1},\ldots,x_{d}) = \overline{ S(x_1,\cdots,x_d) }
\end{align}
where $\overline{S(x_1,\ldots,x_d)} \equiv [S(\overline{x_1}, \ldots, \overline{x_d})]^T$.

As an example, the excitation map for the Majorana plaquette model is given by 
$E(x,y) = (1 + \bar{x} + \bar{y},\,1 + \bar{x} + \bar{x}y )$.
Below, we show the action of the operator $\gamma_a$
at the origin in the Majorana plaquette model,
which creates three adjacent excitations as specified by the red points.
The locations of the excitations are obtained by performing the matrix multiplication of 
$E$ with the polynomial representation $\begin{pmatrix} 1 \\ 0 \end{pmatrix}$ of $\gamma_{a}$:
\begin{align}
E(x,y) \cdot \begin{pmatrix} 1\\ 0 \end{pmatrix} = 1 + \bar{x} + \bar{y}.
\end{align}
Therefore, the action of $\gamma_a$
may be represented by the polynomial $1 + \bar{x} + \bar{y}$,
labeling the locations of the flipped plaquettes;
here, the plaquette operator corresponding to the origin (i.e. the location ``1")
is to the right of $\gamma_a$,
as can be seen from its polynomial representation~\eqref{eq:smap-PlaquetteModel}.

A dictionary that summarizes the relationship between Majorana operators and polynomials is given in Table~\ref{tab:Dictionary}.
\begin{table}
  \begin{tabular}{|c|c|}
  \hline & \\
  {\bf Operator} & {\bf Polynomial}\\ & \\
  \hhline{|=|=|} &\\
  \underline{Majorana Fermion} & \underline{Vector over $\mathbb{F}_{2}$}\\ & \\
  $\begin{array}{c}\gamma_{j} \\ \\ \text{$[j = 1,\ldots,n$ for each}\\ \text{site in the unit cell]} \end{array}$ & $\begin{array}{c}\vec{e}_{j}\\ \\ \text{[$n$-dimensional unit vector }\\ \text{with $j^{\text{th}}$ entry equal to 1]}\end{array}$\\ & \\
    \hline & \\
   \underline{Translation} & \underline{Monomial Multiplication}\\ &\\
   $\boldsymbol{t}_{1}^{n_{1}}\boldsymbol{t}_{2}^{n_{2}}\cdots\boldsymbol{t}_{d}^{n_{d}}\gamma_{j}$ & $x_{1}^{n_{1}}x_{2}^{n_{2}}\cdots x_{d}^{n_{d}}\vec{e}_{j}$\\ & \\
    \hline & \\
   \underline{Multiplication} & \underline{Addition in $\mathbb{F}_{2}[x_{1},\cdots,x_{d}]$}\\ & \\
   $\gamma_{j}\cdot\boldsymbol{t}^{n}_{k}\gamma_{\ell}$ & $\vec{e}_{j} + (x_{k})^{n}\,\vec{e}_{\ell}$\\ & \\
   \hline
  \end{tabular}
\caption{Summary of the polynomial representation of Majorana operators.
An arbitrary operator in $d$ spatial dimensions, written as the product of Majorana fermions,
may be represented as a vector with entries in the (Laurent) polynomial ring 
$\mathbb{F}_{2}[x_1^{\pm 1}, \cdots, x_d^{\pm 1}]$.}
 \label{tab:Dictionary}
\end{table}

\subsection{Topological Order and Ground-State Degeneracy in the Algebraic Representation}

The polynomial representation of Majorana operators 
serves as a starting point for constructing commuting Majorana Hamiltonians that exhibit topological orders.
As we demonstrate in the Supplemental Material~\cite{Supp_Mat},
for a translationally invariant Majorana Hamiltonian with a single operator per lattice site, 
{\it all} operators mutually commute \emph{if and only if} its stabilizer map $S(x_{1},\cdots,x_{d})$ satisfies the condition 
\begin{align}\label{eq:Commutativity}
\overline{S(x_1,\ldots,x_d)}\cdot S(x_{1},\ldots,x_{d}) = 0.
\end{align}
More generally, if the Hamiltonian contains multiple operators per lattice site $\{\mathcal{O}^{(i)}\}$,
then we may define a set of stabilizer maps for each type of operator $\{S_{i}\}$,
so that the condition $\overline{S_{i}(x_{1},\ldots,x_{d})}\cdot S_{j}(x_{1},\ldots,x_{d}) = 0$ for all $i$, $j$,
guarantees that all terms in the Hamiltonian commute.

We next formulate a necessary and sufficient algebraic condition for topological order in commuting Majorana Hamiltonians,
which requires that any degenerate ground-states of a topologically-ordered Hamiltonian 
cannot be distinguished by local operators.
The local indistinguishability is equivalent to the condition that,
for {any} \emph{local} operator $M_{i}$
\begin{align}\label{eq:Local_Indist}
\Pi_{\mathrm{GS}}\,M_{i}\,\Pi_{\mathrm{GS}} = c(M_{i})\,\Pi_{\mathrm{GS}}
\end{align}
where $\Pi_{\mathrm{GS}}$ is the projector onto a ground-state sector
and $c(M_{i})$ is a constant that \emph{only} depends on the operator.
For our case, consider an operator $M_{I}$ that is the product of Majorana operators,
and $P(M_{i})$, the polynomial representation of $M_{i}$.
If $M_{i}$ anti-commutes with \emph{any} term in the Hamiltonian,
then $M_{i}$ creates excitations when acting on the ground-state,
and we have $\Pi_{\mathrm{GS}}\,M_{i}\,\Pi_{\mathrm{GS}} = 0$.
If $M_{i}$ commutes with the Hamiltonian,
then $P(M_{i}) \in \ker E$, as $M_{i}$ creates no excitations.
In this case, the condition $\Pi_{\mathrm{GS}}\,M_{i}\,\Pi_{\mathrm{GS}} = c(M_{i})\Pi_{\mathrm{GS}}$
is guaranteed if $M_{i}$ may be written as a product of operators already appearing in the Hamiltonian.
%{\bf JH: This is not immediate. Revise the logic here.}
More generally, any local operator $M$ that commutes with the Hamiltonian then takes the form:
\begin{align}
M = \sum_{i}M_{i}
\end{align}
where each term $M_{i}$ is the product of operators already appearing in the Hamiltonian.
This condition is necessary for distinct ground-state sectors to be locally indistinguishable.

In our polynomial representation, we enforce the condition \eqref{eq:Local_Indist}
by requiring that the stabilizer and excitation maps satisfy the following condition on an infinite lattice
\begin{align}\label{eq:Top_Order_Condition}
\ker E \cong \mathrm{im}~S.
\end{align}
Recall that the image of $S$ is the set of all polynomial linear combinations of $S(x_1, ..., x_d)$,
taking the form of 
\begin{align}
\sum_{\{ n_i \}} x_1^{n_1} \cdots x_d^{n_d} S( x_1, \ldots, x_d),  
\end{align}
%{\bf JH: mention in the beginning of this section to say that we only consider one interaction term per unit cell.}
and representing all operators that can be written as a product of the commuting operators appearing in the Hamiltonian.
On the other hand, the kernel of the excitation map $E$ is the set of all operators that do not create any excitation when acting on the ground state.  
The above algebraic condition \eqref{eq:Top_Order_Condition} for topological order is thus equivalent to the statement that  
\emph{any} operator that creates no excitations on a ground state on an infinite lattice 
is necessarily a product of operators $\{\mathcal{O}_{n}\}$ already appearing the Hamiltonian.
In other words, there are no %{\bf JH: non-trivial}
non-trivial, locally conserved quantities,
and any degenerate ground-states of the Hamiltonian are locally indistinguishable. In summary, imposing the commutativity (\ref{eq:Commutativity})
and local indistinguishability (\ref{eq:Top_Order_Condition}) conditions on a stabilizer map produces an ideal Majorana Hamiltonian with topological order.
%{\bf JH the commutativity is not derived, but is imposed.
%I think this sentence can be simply removed: }

%. A non-local operator that distinguishes topologically degenerate ground-states on the $d$-dimensional torus $T^{d}$ (e.g. a Wilson line) is an element of $\mathrm{ker}(E)$.  However, any two such operators are equivalent if they can be deformed into each other by acting with the commuting operators appearing in the Hamiltonian.  Therefore, distinct, non-local operators that commute with the Hamiltonian are in one-to-one correspondence with elements of the quotient space $\mathrm{ker}(E)/\mathrm{im}(S)$. However, only half of these operators form a mutually commuting set and may be used to distinguish topologically degenerate ground-states. Therefore, the ground-state degeneracy $D = 2^{k}$ with $2k \equiv \mathrm{dim}_{\mathbb{F}_{2}}[\mathrm{ker}(E)/\mathrm{im}(S)]$. For example, in the fermion parity-graded $Z_{2}$ gauge theory in the Majorana plaquette model, there are two types of Wilson lines, 
%written as a product of Majorana fermions and differing by one sublattice shift \cite{vijay}, for each of the two non-trivial cycles of the torus.  Only half of these Wilson lines -- corresponding to creating and winding electric charges -- mutually commute, with the remaining Wilson lines changing ground-state sectors.  
%For an algebraic calculation, it is convenient to compute the ground-state degeneracy 
%
%
We may compute the ground-state degeneracy of an ideal Majorana Hamiltonian
in the polynomial representation via constraint-counting.
%{\bf JH}{\it fermion parity is already inferred from the eigenvalues of the Hamiltonian terms,
%so we don't really have to constrain the fermion parity. In fact, our degeneracy counting
%includes different fermion parity sectors. So I change this part as the following.
%I think that using Sagar's example of orthogonal matrix over $\mathbb F_2$,
%one can make the following fully rigorous. Though the orthogonal matrix stuff should go into the appendix.}
A lattice with $2M$ Majorana fermions defines a $2^M$-dimensional Hilbert space. On the torus, however, fixing the eigenvalues of the commuting operators in the ideal Majorana Hamiltonian only imposes $M - k$ multiplicatively independent constraints, since the product of certain operators appearing in the Hamiltonian will yield the identity. 
%If there are $M-k$ multiplicatively independent constraints of form 
%$(i) \gamma_{a} \cdots \gamma_{a'} = 1$,
The ground-state degeneracy is simply given by the space of states satisfying the constraints, which is precisely $2^{M}/2^{M-k} = 2^k$.  
%Formally, this can be proved by the fact that there exists a unitary basis change in the Hilbert space 
%such that an even Majorana operator is mapped to a Majorana bilinear.
As each ideal Majorana Hamiltonian in this work consists of exactly one term for each pair of Majorana modes, we see that $k$ is directly equal to the number of constraints on the commuting operators appearing in the Hamiltonian. %In our Majorana cubic model or in the Majorana plaquette model,
%each Hamiltonian term is a constraint for the ground state,
%and we need to know the number of independent terms.
%Since there are exactly one term of the Hamiltonian for a pair of Majorana modes,
%we see that $k$ is directly equal to the number of multiplicative relations among the terms of Hamiltonian.

For example, in the Majorana plaquette model, 
we may group the plaquette operators $\{\mathcal{O}_{p}\}$ 
into three types ($A$, $B$, and $C$) as shown in Figure~\ref{fig:Plaquette_1}(b).
On the torus, the product of the $A$, $B$, and $C$-type operators is identical 
and equal to the total fermion parity~\cite{vijay}.
This yields the following two independent constraints:
\begin{align}\label{eq:Constraints}
\prod_{p\in A}\hat{\mathcal{O}}_{p} \prod_{p\in B}\hat{\mathcal{O}}_{p} 
= \prod_{p\in B}\hat{\mathcal{O}}_{p} \prod_{p\in C}\hat{\mathcal{O}}_{p} = 1,
\end{align}
and produces a $2^2$-fold degenerate ground-state on the torus.  These constraints may be compactly represented using polynomials labeling the locations of the $A$, $B$ and $C$-type plaquettes. 
For example, the collection of all $A$ plaquettes is captured by the polynomial
\begin{align}
 p_A = (1+xy+x^2 y^2)\left( \sum_{n=0}^{L-1} x^{3n} \right)\left( \sum_{m=0}^{L-1} y^{3m} \right).
\end{align}
It is straightforward to expand $p_A$ to verify that the exponents of the non-zero terms describe the positions of $A$ plaquettes.
Here, $L$ specifies the periodic boundary conditions in the $x$ and $y$ directions, so that $x^L=1, ~y^L=1$.
Similarly, the collections of all plaquettes in $B$ and $C$ are encoded in $y p_A$ and $\bar x y p_A$, respectively.
The constraints (\ref{eq:Constraints}) arise from the fact that $(p_A+y p_A) S = 0$, using Eq.~\eqref{eq:smap-PlaquetteModel} and the boundary conditions.

In terms of the stabilizer map, any multiplicative constraint on the operators in the ideal Majorana Hamiltonian on the torus is in one-to-one correspondence with a solution $p$
of the equation $p \cdot S = 0$, so that the polynomial $p$ is an element of the kernel of $S$.
Therefore, the number of independent relations is given by
\begin{align}\label{eq:Degen}
k = \dim_{\mathbb{F}_{2}}[\mathrm{ker}(S)].
\end{align}

We rewrite the expression \eqref{eq:Degen} in a more convenient form for calculations
that will also allow us to make general statements about the scaling behavior of the ground-state degeneracy
with system size for an ideal Majorana Hamiltonian of the form \eqref{eq:Hamiltonian}.
As proven in Corollary 4.5 in Ref.~\cite{Haah_Commuting_Pauli},
Eq.~\eqref{eq:Degen} is equivalent to the dimension of the following quotient ring:
\begin{align}\label{eq:Quotient_Degeneracy}
k = \log_{2} D = \dim_{\mathbb{F}_{2}}\left(\frac{\mathbb{F}_{2}[x_{1},\cdots,x_{d}]}{I(S)+ \mathfrak{b}_{L}}\right).
\end{align}
Here, $I(S)$ the \emph{ideal} generated by the stabilizer map;
if $S^{T} = (s_{1}, \ldots, s_{2n})$ then $I(S)$ is the space of polynomials in $\mathbb{F}_{2}[x_1^{\pm 1},\ldots,x_d^{\pm 1}]$
obtained as a linear combination of $\{ s_i \}$:
\begin{align}
I(S) \equiv \left\{ p = \sum_{i=1}^{2n} c_i s_i \,\,\,\Bigg|\,\,\, c_{i} \in \mathbb{F}_2[x_1^{\pm 1},\ldots,x_d^{\pm1}]\right\}.
\end{align}
We will denote the ideal generated by a set $\{s_{1},\ldots, s_{n}\}$
by $\langle s_1,\ldots, s_{n} \rangle$.
Furthermore, we define the ideal $\mathfrak{b}_L \equiv \langle x_1^L - 1, \ldots, x^L_d - 1\rangle$.
As the quotient space identifies the zero element in $\mathbb{F}_{2}[x_1^{\pm1},\ldots,x_d^{\pm1}]$
with the generators of $I(S) + \mathfrak{b}_{L}$,
we observe that the ideal $\mathfrak{b}_{L}$ is used to enforce
the periodic boundary conditions on a $d$-dimensional torus with side-length $L$.  

We emphasize that the ideal $I(S)$ is the space of excitations
that can be created through the action of any operator on the ground-state.
Therefore, the expression \eqref{eq:Quotient_Degeneracy}
may be physically interpreted as counting certain superselection sectors of the ideal Majorana Hamiltonian. 
Any $p\in\mathbb{F}_{2}[x_1^{\pm1},\cdots,x_d^{\pm1}]$
corresponds to a virtual %{\bf JH: (a virtual)}
eigenstate of the Hamiltonian with excitations at the locations specified by the polynomial $p$.
Certain states, however, cannot be created by acting with an operator on a ground-state $\ket{\Psi}$
due to the $k$ constraints on the commuting operators.
For example, in the Majorana plaquette model,
it is impossible to obtain a state with a single plaquette excitation by acting on the ground-state,
since the products of $A$, $B$ and $C$ plaquettes must satisfy \eqref{eq:Constraints}.
As $I(S) / (I(S) \cap \mathfrak{b}_{L})$ is the set of excitations
that can be created by the action of operators on the ground-state for a finite system,
the quotient space 
$\left( \mathbb{F}_2[x_1^{\pm1},\cdots, x_d^{\pm1}]/\mathfrak b_L \right)/ \left( I(S)/(I(S)\cap\mathfrak{b}_{L}) \right)
= \mathbb{F}_2[x_1,\cdots, x_d] / \left( \mathfrak b_L + I(S) \right)$
is the set of virtual %{\bf JH: virtual}
eigenstates of the Hamiltonian that cannot be deformed
into each other through the action of any local operator. %{\bf JH: removed ``translation''}
For the Majorana plaquette model, this quotient space is 
\begin{align}
\frac{\mathbb{F}_{2}[x,y]}{\langle 1 + x + y,\,x + y + xy,\,x^{L} - 1,\,y^{L}-1\rangle} \cong \mathbb F_2^2
\end{align}
when $L$ mod $3 = 0$ so that there are an equal number of $A$, $B$, and $C$ plaquettes.
In this case, the trivial vacuum $(0)$ and a state with a single plaquette excitation $(1)$ on $A$, $B$, or $C$
correspond to the four superselection sectors in the quotient ring.

%The space of all possible excitations that \emph{can} be treated by acting on the ground-state is given by considering $I(S)$, the \emph{ideal} generated by the stabilizer map; if $S^{T} = (s_{1}, \ldots, s_{2n})$ then $I(S)$ is the space of polynomials over $\mathbb{F}_{2}$ obtained as a linear combination of $\{s_{i}\}$, i.e.
%\begin{align}
%I(S) = \left\{p = \sum_{i=1}^{2n}c_{i}s_{i}\,\,\,\Bigg|\,\,\, c_{i} \in \mathbb{F}_{2}[x_{1},\ldots,x_{d}]\right\}.
%\end{align}
%We will denote the ideal generated by a set $\{s_{1},\ldots, s_{2n}\}$ by $\langle s_{1},\ldots, s_{2n} \rangle$. Similarly, let $\mathfrak{b}_{L} = \langle x_{1}^{L} - 1, \ldots, x^{L}_{d} - 1\rangle$ be an ideal that enforces the boundary conditions on a $d$-dimensional torus with side-length $L$.  The combined ideal $I(S)\oplus\mathfrak{b}_{L}$ forms a natural equivalence relation for Majorana operators on the torus due to the action of the commuting operators appearing in the Hamiltonian. Distinct elements of the quotient ring $\mathbb{F}_{2}[x_{1},\ldots,x_{d}]/(I(S)\oplus\mathfrak{b}_{L})$ correspond to operators that can be deformed into each other by acting with the Hamiltonian. The ground-state degeneracy $D$ is therefore computed \cite{Haah_Commuting_Pauli} from the dimension of this space as

\begin{table}
  \begin{tabular}{|c|c|}
  \hline & \\
  {\bf Algebraic Expression} & {\bf \,\,\,Physical Interpretation\,\,\,}\\ & \\
  \hhline{|=|=|} & \\
  $\begin{array}{c}\overline{S(x_{1},..,x_{d})}\cdot S(x_{1},..,x_{d}) = 0 \\\end{array}$ & \\ & \parbox[c]{4cm}{\vspace{-1.1\baselineskip}\raggedright Commutativity condition, that all operators $\{\mathcal{O}_{n}\}$ appearing in the Hamiltonian mutually commute.}  \\ & \\
  \hline & \\
$\begin{array}{c}\mathrm{im}(S)\\\end{array}$ & \\ & \parbox[c]{4cm}{\vspace{-1.1\baselineskip}\raggedright Set of operators that may be written as the product of  commuting operators $\{\mathcal{O}_{n}\}$ in the Hamiltonian.}  \\ & \\
    \hline & \\
   $\begin{array}{c}\mathrm{ker}(E)\\\end{array}$ & \\ & \parbox[c]{4cm}{\vspace{-1.1\baselineskip}\raggedright Set of operators that create no excitations when acting on the ground-state $\ket{\Psi}$.} \\ & \\
   \hline &\\
   $\begin{array}{c} k = \mathrm{dim}_{\mathbb{F}_{2}}[\mathrm{ker}(S)]\\\end{array}$ & \\ & \parbox[c]{4cm}{\vspace{-1.1\baselineskip}\raggedright The number of independent relations among the commuting operators in the Hamiltonian, when placed on a torus. The ground state degeneracy $D = 2^{k}$.} \\ & \\
    \hline & \\
   $\begin{array}{c} p\in\mathbb{F}_{2}[x_{1}^{\pm},\ldots,x_{d}^{\pm}]\\\end{array}$ & \\ & \parbox[c]{3.8cm}{\vspace{-1.1\baselineskip}\raggedright A configuration of excitations, specified by the locations of operators $\{\mathcal{O}_{n}\}$ with flipped eigenvalue $-1$.} \\ & \\
   \hline & \\
   $\begin{array}{c} q \in I(S)\\\end{array}$ & \\ & \parbox[c]{4cm}{\vspace{-1.1\baselineskip}\raggedright A configuration of excitations that may be created by acting with an operator on the ground-state $\ket{\Psi}$.}\\ & \\
   \hline 
  \end{tabular}
\caption{Dictionary of various algebraic quantities and their physical interpretation in the context of a commuting Majorana Hamiltonian. \label{tab:Alg_Dictionary}}
\end{table}

\begin{table*}
\begin{tabular}{|c|c|c|c|c|c|c|c|}
  \hline
  & $f_{1}(x,y,z)$ & $f_{2}(x,y,z)$ & $f_{3}(x,y,z)$ & $f_{4}(x,y,z)$ & $f_{5}(x,y,z)$ & $f_{6}(x,y,z)$\\
  \hline
  \parbox[c][.5in]{0.25in}{$f$} & $ 1 + x + y + z $ & $ \begin{array}{c} 1 + z + xy\\ + yz + xz\end{array}$ & $ \begin{array}{c} 1 + x + y\\ + yz + xz\end{array}$ & $\begin{array}{c} 1 + y + z\\ + xy + yz + xz\end{array}$ & $\begin{array}{c}1 + x + y + z\\ + xy + yz + xz\end{array}$ & $\begin{array}{c}1 + x + y\\ + z + yz\end{array}$\\
  \hline
  $\begin{array}{c}\\\mathcal{O}\end{array}$ &  
   \parbox[c][1.1in]{1in}{\includegraphics[trim = 170 390 600 50, clip = true, width=0.11\textwidth, angle = 0.]{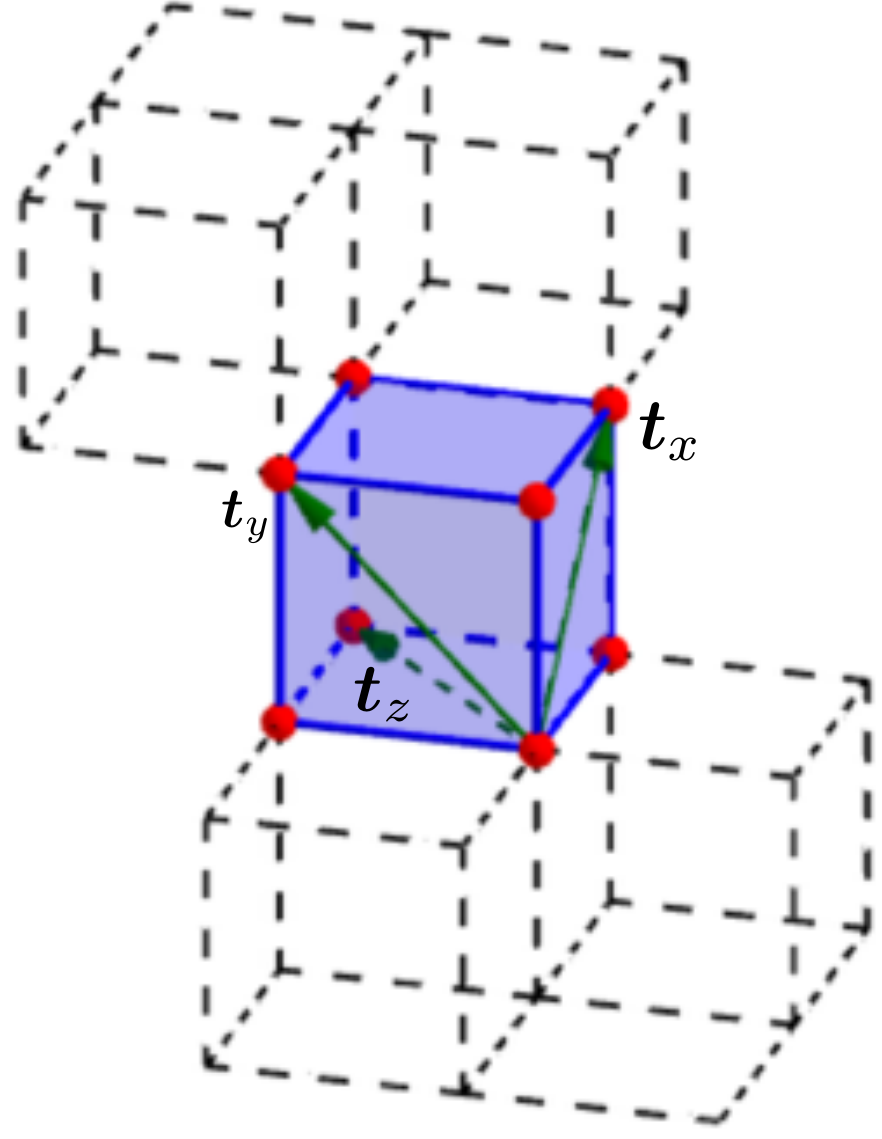}}
   & 
   \parbox[c][1.1in]{1in}{\includegraphics[trim = 180 390 600 50, clip = true, width=0.11\textwidth, angle = 0.]{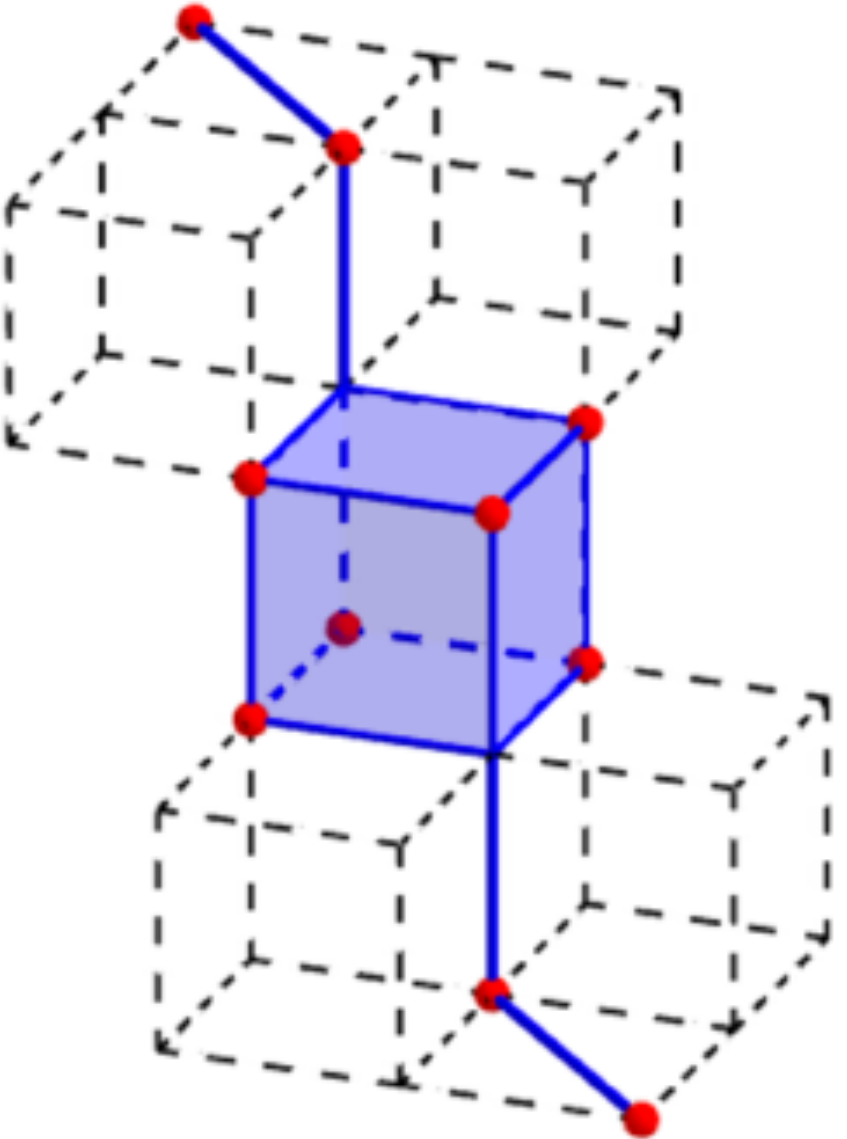}} 
   &
   \parbox[c][1.1in]{1in}{\includegraphics[trim = 180 390 600 50, clip = true, width=0.11\textwidth, angle = 0.]{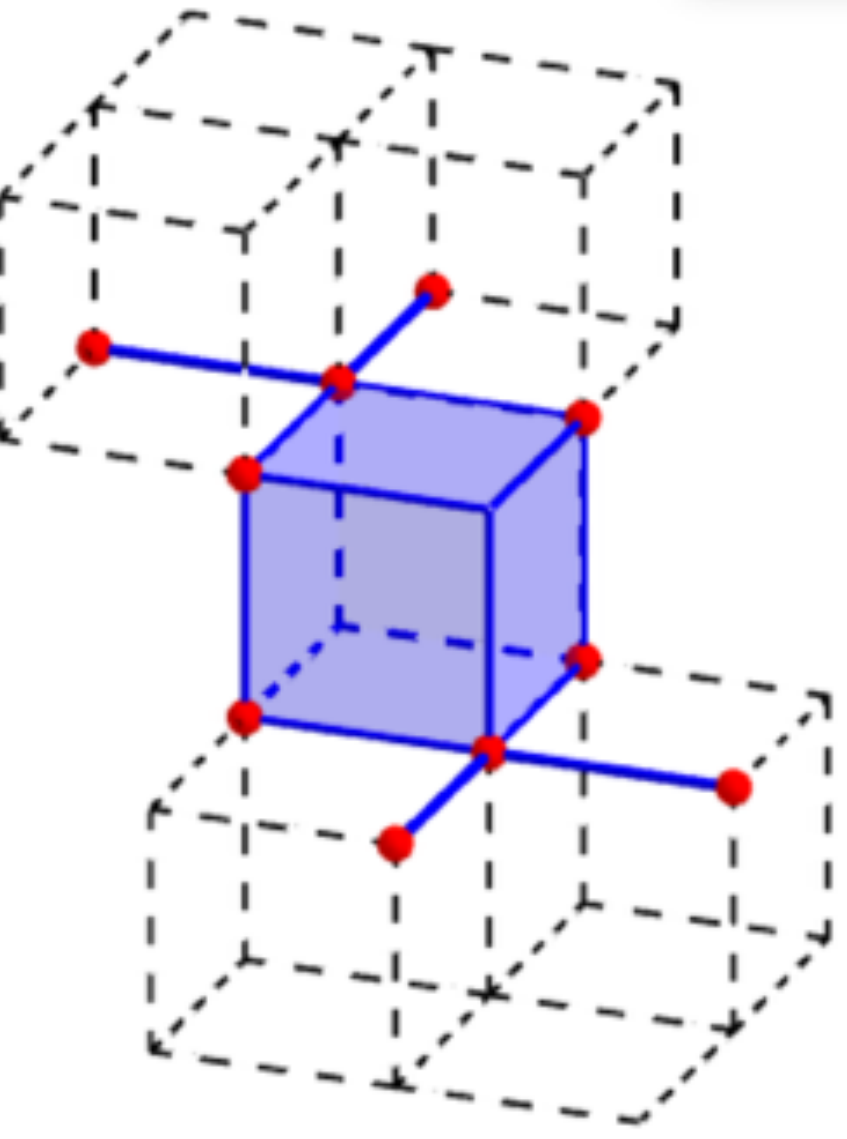}} 
   &
   \parbox[c][1.1in]{1in}{\includegraphics[trim = 180 390 600 50, clip = true, width=0.11\textwidth, angle = 0.]{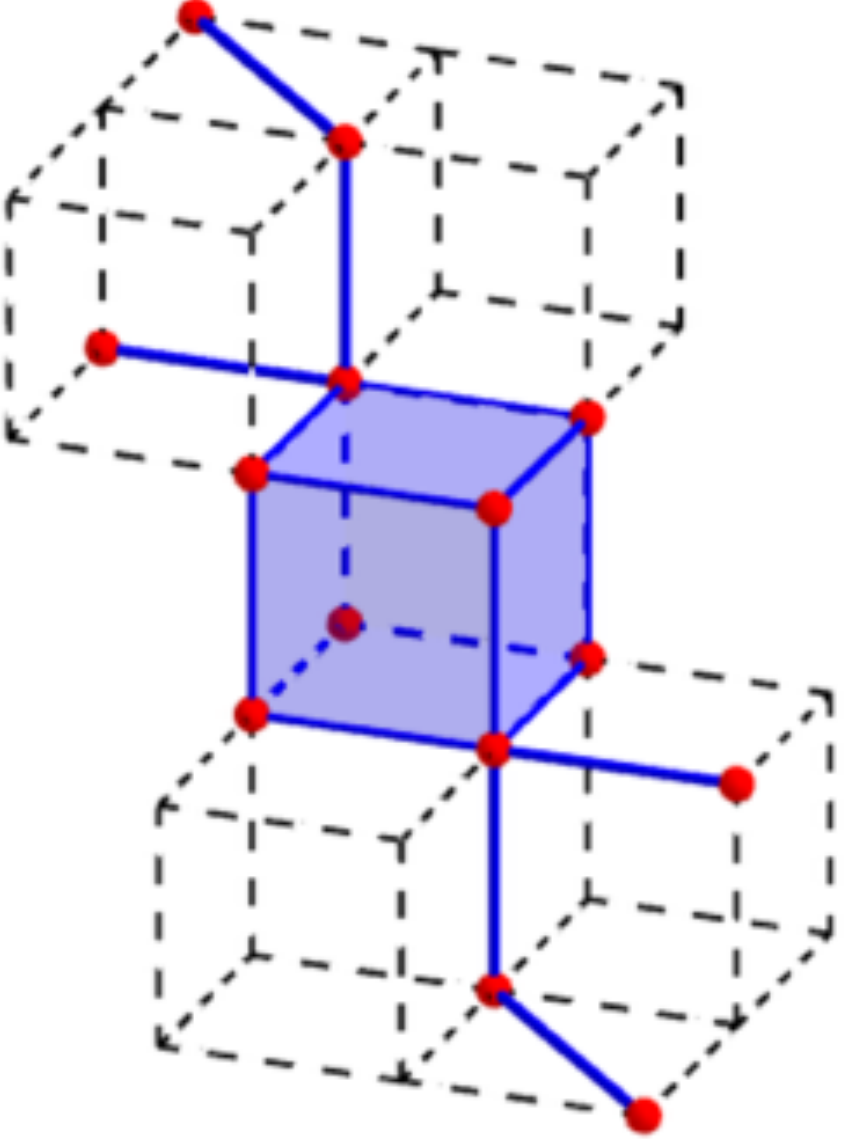}}
   & 
   \parbox[c][1.1in]{1in}{\includegraphics[trim = 180 390 600 50, clip = true, width=0.11\textwidth, angle = 0.]{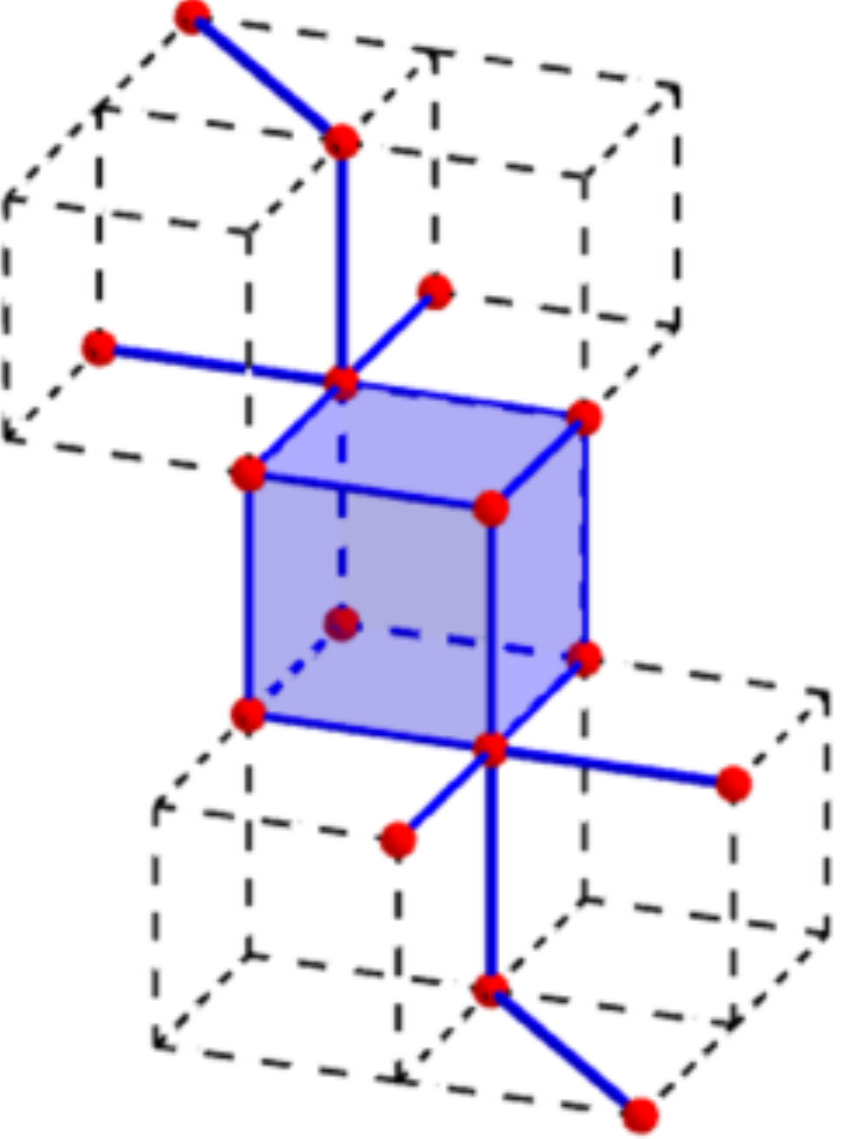}}
   & 
   \parbox[c][1.1in]{1in}{\includegraphics[trim = 180 390 600 50, clip = true, width=0.11\textwidth, angle = 0.]{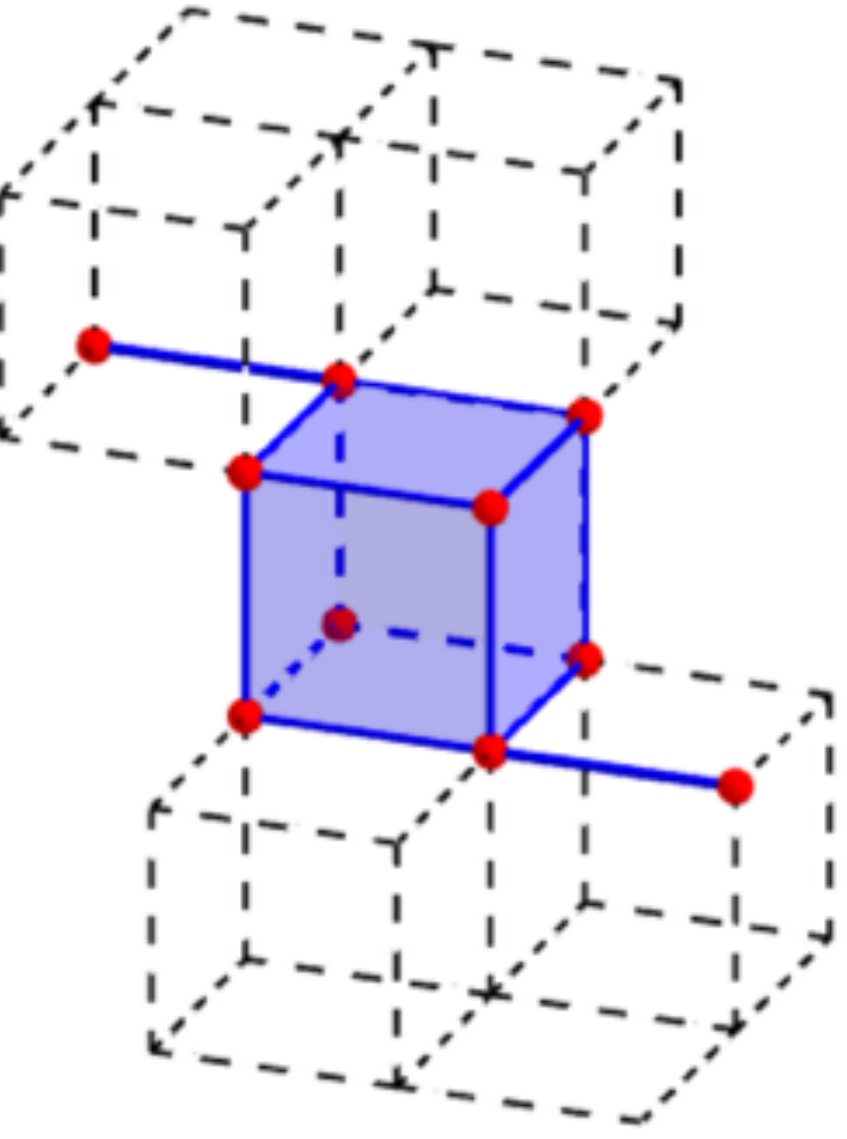}}\\
   \hline
   $\log_{2}D$  
   & $3L-2$
   & $\begin{array}{c}6L - 6\,\, (L = 3n)\\0\,\,(L \ne 3n)\end{array}$
    & $\begin{array}{c}\\4L/3\,\,(L = 6\cdot 2^{n})\\8L/5\,\,(L = 5\cdot 2^{n})\\\vdots\\\end{array}$
   & $\begin{array}{c}4L - 4\,\,(L = 2n)\\2L - 1\,\,(L = 2n+1)\end{array}$ 
   & $\begin{array}{c}\\2L - 2\,\,(L = 2^{2n+1} - 1)\\2L - 4\,\,(L = 2^{2n} - 1)\\\vdots\\\end{array}$
   & $\begin{array}{c}\\4L/3\,\,(L = 6\cdot 2^{n})\\8L/5\,\,(L = 5\cdot 2^{n})\\\vdots\\\end{array}$\\
   \hline
   \end{tabular}
 \caption{We find $7$ distinct, topologically-ordered ideal Majorana Hamiltonians
 with nearest-neighbor interactions on a lattice with a two-site unit cell in $d = 3$ spatial dimensions.
 The first model $f_{0}(x,y,z) = 1 + y + z$ (not shown) is a trivial stack of two-dimensional Majorana plaquette models, considered in Ref.~\cite{vijay}.
 For the remaining $6$ models, the action of the elementary operator $\mathcal{O}$
 appearing in the ideal Majorana Hamiltonian is shown above
 as the product of the Majorana fermions on the indicated red dots.
 In the depiction of the Majorana cubic model $f_{1}(x,y,z)$, we have also shown the choice of translation vectors $\{\boldsymbol{t}_{x},\boldsymbol{t}_{y},\boldsymbol{t}_{z}\}$ on the lattice,
 originating from one of the sites within the unit cell; to compute the ground-state degeneracy on an $L\times L\times L$ torus, we impose periodic boundary conditions by requiring that $\boldsymbol{t}_{x}^{L} = \boldsymbol{t}_{y}^{L} = \boldsymbol{t}_{z}^{L} = 1$.  The topological ground-state degeneracy ($D$) of each of these models is extensive.  For models $f_{3}(x,y,z)$, $f_{5}(x,y,z)$, and $f_{6}(x,y,z)$, the  
ground-state degeneracy on the three-torus is a highly sensitive function of system size, and only the maximum value of the degeneracy is shown for the indicated choices of $L$.
  }
 \label{table:3D_Majorana_Models}
\end{table*}

The expression for the ground-state degeneracy \eqref{eq:Quotient_Degeneracy}
is convenient as the dimension of a quotient ring may be computed using algebraic techniques.
Most often, we will determine a Gr\"{o}bner basis for the ideal $I(S)+\mathfrak{b}_{L}$
in order to determine membership in the quotient ring.
For a polynomial ring $R$, we may define a total monomial ordering
(e.g. lexicographic order with $x_{1} \succ x_{2} \succ \ldots \succ x_{d}$);
we denote the leading monomial in a polynomial $h\in R$ as $\mathrm{LM}(h)$ with respect to this ordering.
Given an ideal $I = \langle s_{1}, \cdots, s_{n}\rangle$ of a polynomial ring,
there exists a canonical choice of generators for the ideal,
known as the Gr\"{o}bner basis $\{ g_{1}, \cdots, g_{n} \}$,
with the property that for any $f \in I$,
$\mathrm{LM}(f) \in \langle \mathrm{LM}(g_{1}),\ldots, \mathrm{LM}(g_{n})\rangle$,
i.e. any element of the ideal has a leading term contained in the ideal generated by the leading terms of the Gr\"{o}bner basis.
As a result, the dimension of the quotient ring $\mathrm{dim}[R/I]$
is merely given by the number of monomials that are smaller (in the monomial ordering) 
than all of the leading terms in the Gr\"{o}bner basis.
This is because any polynomial $p\in R$ may be reduced by the Gr\"{o}bner basis
until the leading term of the reduced polynomial satisfies $\mathrm{LM}(p_{\mathrm{red}}) < \mathrm{LM}(g_{i})$ for all $i = 1,\ldots,n$.
Therefore, each monomial $m$ satisfying $m < \mathrm{LM}(g_{i})$ for all $i$
corresponds to a unique representative of the quotient ring $R/I$.

We note that calculations of the ground-state degeneracy for any commuting Majorana Hamiltonians presented in this work are done by determining a Gr\"{o}bner basis for the ideal $I(S) + \mathfrak{b}_{L}$. In this way, the calculation of the degeneracy reduces to  counting points in an algebraic set.
%In the Supplemental Material~\cite{Supp_Mat},
%we apply this technique to the Majorana plaquette model.
%The Gr\"{o}bner basis for the ideal $I(S) + \langle x^{L} - 1, y^{L} - 1\rangle$
%is calculated and used to demonstrate 
%that the ground-state of the Majorana plaquette model is indeed four-fold degenerate on 2-torus.

\subsection{Unitary and Stable Equivalence}
The polynomial representation of the ideal Majorana Hamiltonian contains built-in redundancies,
since we may re-define the unit cell or translation operators on the $d$-dimensional lattice.
For the stabilizer map, the translation corresponds to multiplication of any entry of $S(x_{1},\ldots,x_{d})$ by a monomial. 
In this way, a stabilizer map $S(x_{1},\ldots,x_{d})$ is only defined up to monomial multiplication on each of its entries.
Furthermore, for an ideal Majorana Hamiltonian with longer-range interactions,
we may always enlarge the unit cell.
As our focus will be on Majorana models with nearest-neighbor interactions,
we neglect this redundancy in the stabilizer map.

Equivalence relations, given by local unitary transformations on ideal Majorana Hamiltonians,
may also be considered in the polynomial language.
For instance, two ideal Majorana Hamiltonians,
defined by stabilizer maps $S(x_1,\ldots,x_d)$ and $S'(x_1,\ldots,x_d)$
are \emph{unitarily} equivalent if there exists a matrix $U$
such that $S'(x_1,\ldots,x_d) = U \cdot S(x_1,\ldots,x_d)$
where $U \in O(n; \mathbb{F}_{2})$, an orthogonal matrix over $\mathbb{F}_{2}$
satisfying $U^T U = 1$.
This guarantees that if $\overline{S(x_{1},\ldots,x_{d})} \cdot S(x_{1},\ldots,x_{d}) = 0$,
then $\overline{S'(x_{1},\ldots,x_{d})}\cdot S'(x_{1},\ldots,x_{d}) = 0$ as well.
Finally, we take two stabilizer maps to be \emph{stably} equivalent
if we can obtain one from the other by attaching a trivial (dimerized) set of Majorana fermions.
This is expressed as $S(x_{1},\ldots,x_{d})^{T} \sim S(x_{1},\ldots,x_{d})^{T} \oplus \left(0,\cdots,0,1,1\right)$.

\section{Extensive Topological Degeneracy in $d \ge 3$}
\label{sec:extensive}

Using the commutativity \eqref{eq:Commutativity} and local indistinguishability \eqref{eq:Top_Order_Condition} conditions,
and the built-in redundancy in the polynomial description,
we demonstrate in the Supplemental Material \cite{Supp_Mat},
that an ideal Majorana Hamiltonian defined on a $d$-dimensional lattice with a two-site basis is topologically-ordered
if the stabilizer map may be written in the following form, 
after multiplying each entry by appropriate monomials:
\begin{align}\label{eq:Two_Site_Stabilizer}
S = \begin{pmatrix}
f(x_{1},\cdots, x_{d})\\\\
 \overline{f(x_{1},\cdots, x_{d})}
\end{pmatrix}
\end{align}
where $f(x_1,\cdots, x_d)\in\mathbb{F}_{2}[x_1^{\pm1},\cdots,x_d^{\pm1}]$ and $f$ and $\bar{f}$ are co-prime,
i.e., $f$ and $\bar{f}$ have no common polynomial factors.
As a result, a topologically-ordered, ideal Majorana Hamiltonian
with a two-site basis may be specified by a single polynomial.
For example, the stabilizer map for the Majorana plaquette model
takes the form $S^{T} = (f(x,y), \,x\cdot \overline{f(x,y)}\,)$ with $f(x,y) = 1 + x + y$.  

The dimension of the quotient ring \eqref{eq:Quotient_Degeneracy}
scales as the dimension of the space of the zeros of the ideal $I(S)$ over the field extension $\mathbb{F}_{2^m}$
when $L = 2^m -1$.
As a result, for an ideal Majorana Hamiltonian \eqref{eq:Hamiltonian}
with a two-site unit cell, the space of solutions to 
\begin{align}
f(x_{1},\ldots,x_{d}) = 0, \quad \quad \overline{f(x_{1},\ldots,x_{d})} = 0
\end{align}
generally defines an $(d - 2)$-dimensional variety,
so that the ground-state degeneracy scales on the $d$-dimensional torus with side-length $L$
as $\mathrm{log}_{2}D = cL^{d-2} + \cdots$ for some constant $c$.
We emphasize that this produces a class of ideal Majorana models with extensive topological degeneracy in $d = 3$ dimensions.
%{\bf JH: This is also ``generic'' for Pauli case: whereas this is a highly non-generic feature of three-dimensional commuting Pauli Hamiltonians.  }
%{\bf JH: I would add}
Remarkably, while our models have a two-dimensional Hilbert space and a single interaction term per lattice site, this only constrains the full Hilbert space up to extensive topological degeneracy.

We have exhaustively searched for distinct,
ideal Majorana Hamiltonians with a two-site basis and nearest-neighbor interactions in $d = 2$ and $d = 3$ spatial dimensions.
This is straightforward as the orthogonal group $O(2; \mathbb{F}_{2}) = \{\mathds{1}_{2\times 2}, \,\sigma^{x}\} $
so that the space of local unitary transformations between these ideal Majorana Hamiltonians is trivial.
In $d = 2$ spatial dimensions, the only such Hamiltonian is the Majorana plaquette model with
\begin{align}
f(x,y) = 1 + x + y.
\end{align}

In $d = 3$ dimensions, however, we find $7$ distinct Majorana models
with a two-site basis and nearest-neighbor interactions.  The first model has the polynomial representation $f_{0}(x,y,z) = 1 + y + z$ and is a trivial stack of two-dimensional Majorana plaquette models. 
The polynomial representations of the remaining models,
along with their ground-state degeneracies on a torus of side-length $L$ are shown in Table~\ref{table:3D_Majorana_Models}.  For simplicity, we have imposed periodic boundary conditions by requiring that $\boldsymbol{t}_{x}^{L} = \boldsymbol{t}_{y}^{L} = \boldsymbol{t}_{z}^{L} = 1$ for the translation vectors $\{\boldsymbol{t}_{x}, \boldsymbol{t}_{y}, \boldsymbol{t}_{z}\}$ shown in the representation of the Majorana cubic model $f_{1}(x,y,z) = 1 + x + y + z$ in Table ~\ref{table:3D_Majorana_Models}. 
Each of the models shown exhibits extensive topological degeneracy and admits at least one topological excitation that is free to move in a sub-manifold of the full lattice.

\section{Fracton Excitations and Dimension-$n$ Anyons}
\label{sec:fracton}

A remarkable feature of these Majorana models is the presence of fundamental excitations that are either perfectly immobile or only free to move in a sub-manifold of the lattice;
attempting to move these excitations by acting with any local operator
will necessarily create additional excitations.
A bound-state of these immobile excitations, however,
forms a particle that can freely move along a higher-dimensional sub-manifold.  

The existence of a fracton fundamental excitation
may be shown rigorously in the polynomial representation of the Majorana models.
An element $p\in I(S)$ of the ideal defined by the stabilizer map
corresponds to a set of excitations that may be created by acting on the ground-state.
The fundamental excitation is mobile
if and only if it is possible to create an isolated pair of such excitations.
Therefore, an ideal Majorana model admits fracton excitations if the stabilizer ideal contains no binomial terms, i.e.
\begin{align}
1 + x_{1}^{n_{1}}x_{2}^{n_{2}}\cdots x_{d}^{n_{d}} \notin I(S)
\end{align}
for any $n_{i}\in\mathbb{Z}$.
%{\bf JH: this is not a concern. If there is a binomial $1+b$ caused by $m$, a polynomial for Majorana operator,
%then $1+b^2 = 1+b +b + b^2$ can be caused by $m+ bm$.
%The parity of $m+bm$ is always even no matter what the parity of $m$ is.}
%We note that even if $I(S)$ contains binomial terms,
%it must be verified that the pair of fundamental excitations
%may be created with a physical operator involving an even number of Majorana fermions.

We now apply the polynomial criterion for fracton excitations
to the Majorana cubic model and to the model $f_{5}(x,y,z) = 1 + x + y + z + xy + yz + xz$,
both shown in Table~\ref{table:3D_Majorana_Models}.

\subsection{Fractons in the Majorana Cubic Model}
We consider the Majorana cubic model,
specified by the single polynomial  $f(x,y,z) = 1 + x + y + z$,
so that the stabilizer map is given by $S = \left(f(x,y,z),\,\overline{f(x,y,z)}\right)^{T}$.
We wish to prove that the ideal generated by the stabilizer map $I(S)$
contains no binomial terms, so that the fundamental cube excitation is a fracton.
This may be shown by considering the zero-locus of the ideal,
i.e., the solutions to the zeros of the generators of the ideal:
\begin{align}\label{eq:Zero_Locus}
1 + x + y + z &= 0\\
xyz + xy + yz + xz &= 0.
\end{align}
A polynomial $p$ belong to $I(S)$ only if %{\bf( JH not ``if'')}
$p$ vanishes on the zero-locus of the ideal.
Note that solutions to \eqref{eq:Zero_Locus} take the form
$(x,y,z) = (1,\alpha,\alpha)$, $(\alpha,1,\alpha)$ or $(\alpha,\alpha,1)$,
where $\alpha$ is an arbitrary element in the extension of $\mathbb F_2$.
However, we see that the binomial $1 + x^{n}y^{m}z^{\ell}$ vanishes on this space of solutions
only if $n = m = \ell = 0$, 
in which case the binomial is zero.
Therefore, we conclude that 
\begin{align}
1 + x^{n}y^{m}z^{\ell}\notin I(S).
\end{align}
As a result, there is no way to create the fundamental cube excitation in the Majorana cubic model in pairs.
Therefore, the cube excitation is an immobile fracton;
a single cube excitation cannot be moved without creating additional excitations.

\subsection{Dimension-1 Fundamental Excitations in $f_{5}(x,y,z)$}
Now, we consider the isotropic model $f_{5}(x,y,z) = 1 + x + y + z + xy + yz + xz$,
with stabilizer map defined by $S(x,y,z) = \left(f_{5}(x,y,z),\,xyz\cdot\overline{f_{5}(x,y,z)}\right)$.
From the excitation map $E(x,y,z) \equiv \overline{S(x,y,z)}$,
we find that the Majorana bilinear along the (1,1,1) direction creates a pair of fundamental excitations:
\begin{align}\label{eq:111_String}
E(x,y,z)\cdot\left(\begin{array}{c} 1\\ 1\end{array}\right) = 1 + \overline{xyz}.
\end{align}  
Therefore, the fundamental excitation in this model is clearly not a fracton.  We now demonstrate that the fundamental excitation may \emph{only} hop freely along the (1,1,1) direction, without creating additional excitations. Consider the variety $V(I)$ defined by the stabilizer ideal $I(S) = \langle 1 + x + y + z + xy + yz + xz,\,xyz + x + y + z + xy + yz + xz \rangle $, i.e. the zero-locus of the generators of the ideal over an extension of $\mathbb{F}_{2}$.  The following is a point on the variety:
\begin{align}\label{eq:f5_variety}
(x,y,z) = \left(t,\,\frac{1}{1+t},\,\frac{t+1}{t} \right)
\end{align}
with $t$ in an extension of $\mathbb{F}_{2}$.  As a result, if $1 + x^{n}y^{m}\in I(S)$, we must have from (\ref{eq:f5_variety}) that $t^{n} = (1 + t)^{m}$ for infinitely many $t$.  This can only be true if $n = m = 0$. As a result, the fundamental excitation cannot hop freely in the $xy$-plane.  As the generators of the ideal are symmetric under exchanging any pair of variables (e.g. $x\longleftrightarrow y$), we conclude that $1 + y^{n}z^{m}$, $1 + x^{n}z^{m}\in I(S)$ only if $n = m = 0$, so that the fundamental excitation cannot freely hop in the $yz$- or $xz$-planes.  From these results, we are led to the conclusion that
\begin{align}
1 + x^{n}y^{m}z^{\ell}\notin I(S)
\end{align}
when $n$, $m$ and $\ell$ are distinct.  Therefore, we have shown that the fundamental excitation in the model defined by $f_{5}(x,y,z)$ is restricted to hop along the (1,1,1) direction of the cubic lattice without creating additional excitations.

\appendix

\section{Stabilizer \& Excitation Maps in Polynomial Representations of Majorana Hamiltonians}
\label{app:commutingOperators}

In this section, we prove a statement presented in the main text, that in a commuting Majorana Hamiltonian, of the general form
\begin{align}
H = -\sum_{m}\left[\mathcal{O}^{(1)}_{m} + \mathcal{O}^{(2)}_{m} + \cdots + \mathcal{O}^{(N)}_{m}\right]
\end{align}
two Majorana operators $\mathcal{O}^{(i)}_{n}$ and $\mathcal{O}^{(j)}_{m}$ commute over all lattice sites (i.e. $[\mathcal{O}^{(i)}_{n}, \mathcal{O}^{(j)}_{m}] = 0$ for any $n$, $m$) if and only if their respective stabilizer maps $S_{i}$ and $S_{j}$ satisfy $\bar{S}_{i}\cdot S_{j} = 0$.  Furthermore, we show that the excitation map $E_{i} = \bar{S}_{i}$ may be constructed for each operator in the Hamiltonian, so that, given the polynomial representation $P$ of some operator, the quantity $E_{i}\cdot P$ yields a polynomial encoding the pattern of excitations created by that operator when acting on the ground-state.  

Consider a $d$-dimensional lattice with translation group $\Lambda \cong \mathbb{Z}^{d}$, and a single Majorana fermion per lattice site.  At a given site, the identity operator $\mathds{1}$ and the Majorana fermion $\gamma$ form the group $\mathbb{Z}_{2}\cong \mathbb{Z}/2\mathbb{Z}$ under multiplication since 
\begin{align}
&\mathds{1} \times \gamma = \gamma \times \mathds{1} = \gamma\\
&\mathds{1} \times \mathds{1} = \gamma \times \gamma = \mathds{1}.
\end{align}
We refer to this as the `Majorana group' $\mathcal{M}$ at a given lattice site.  The group $\mathcal{M}$ naturally forms a vector space over the finite field $\mathbb{F}_{2}$ since for some $\alpha\in\mathbb{Z}_{2}$ and $m\in \mathcal{M}$ we may define the $\mathbb{Z}_{2}$ action $\alpha\cdot m \equiv m^{\alpha} \in \mathcal{M}$.  Furthermore, any element of the translation group $g\in\Lambda$ has a natural action on $\mathcal{M}$ by multiplication.  For instance, let $c, d\in\mathbb{Z}_{2}$, so that we may represent the action of $g$ on the operator $(\gamma)^{c}$ simply as $g\cdot c\in\mathbb{F}_{2}$. A more general operator, for instance $(g\cdot \gamma)^{c}\otimes(h\cdot\gamma)^{d}$ may be written as $g\cdot c + h\cdot d\in\mathbb{F}_{2}$.

Now, we consider a $d$-dimensional lattice with an $n$-site basis so that each site again contains a single Majorana fermion; $n$ is restricted to be an \emph{even} integer so that we may have a well-defined number of \emph{complex} fermions per lattice site. We represent each of the $n$ Majorana fermions at a given lattice site as $\gamma_{j}$, with $j = 1$, $\ldots$, $n$.  Recall that distinct Majorana fermions anti-commute.  Let $g\in\Lambda\cong\mathbb{Z}^{d}$ be an element of the lattice translation group.  From the natural action of the translation group on the Majorana fermions, we may write any two Hermitian operators $\mathcal{O}_{1}$ and $\mathcal{O}_{2}$ up to overall factors of $\pm i$ as
\begin{align}
&\mathcal{O}_{1} \equiv \prod_{g\in\Lambda}\left[\prod_{j = 1}^{n}\left(g\cdot\gamma_{j}\right)^{c_{j}(g)}\right]\\
&\mathcal{O}_{2} \equiv \prod_{g\in\Lambda}\left[\prod_{j = 1}^{n}\left(g\cdot\gamma_{j}\right)^{d_{j}(g)}\right]
\end{align}
with the coefficients $c_{j}(g)$, $d_{j}(g)\in\mathbb{Z}_{2}$.  On physical grounds, we require that each operator is the product of an \emph{even} number of Majorana fermions so that the total fermion parity is conserved.  This is equivalent to the condition that
\begin{align}
\sum_{g\in\Lambda}\sum_{j = 1}^{n}c_{j}(g) = \sum_{g\in\Lambda}\sum_{j = 1}^{n}d_{j}(g) = 0 \in\mathbb{F}_{2}
\end{align}

Two operators commuting with the total fermion parity commute if and only if they have overlapping support on an even number of Majorana fermions.  As a result, the operators $\mathcal{O}_{1}$ and $h\cdot\mathcal{O}_{2}$ commute for some $h\in\Lambda$ if
\begin{align}\label{eq:Commuting_Condition}
\sum_{g\in\Lambda}\sum_{j = 1}^{n}c_{j}(g)\, d_{j}(h^{-1}g) = 0.
\end{align}

The condition (\ref{eq:Commuting_Condition}) may be simply imposed by representing the operators $\mathcal{O}_{1}$ and $\mathcal{O}_{2}$ as
\begin{align}
S_{1} \equiv \sum_{g\in\Lambda}g\cdot\left(\begin{array}{c}
c_{1}(g)\\
c_{2}(g)\\
\vdots\\
c_{n}(g)
\end{array}
\right) \hspace{.125in}
S_{2} \equiv \sum_{g\in\Lambda}g\cdot \left(\begin{array}{c}
d_{1}(g)\\
d_{2}(g)\\
\vdots\\
d_{n}(g)
\end{array}
\right)
\end{align}
respectively.  We further define $\bar{\sigma}_{i}$ by combining the antipode map $g \rightarrow g^{-1}$ with transposition of the $S_{i}$ vector.  For example, 
\begin{align}
\overline{{S}_{1}} = \sum_{g\in\Lambda}g^{-1}\cdot\Big( \begin{array}{cccc}
c_{1}(g) & c_{2}(g) & \cdots & c_{n}(g)
\end{array}\Big)
\end{align}
Now, we demonstrate the following statement: $\overline{{S}_{2}}\cdot S_{1} = 0$ if and only if $[\mathcal{O}_{1},\, \ell\cdot\mathcal{O}_{2}] = 0$ $\forall$ $\ell\in\Lambda$.

We demonstrate this by explicit calculation.  First note that:
\begin{align}
\overline{S_{2}}\cdot S_{1} &= \sum_{g, h\in\Lambda}\sum_{j = 1}^{n}h^{-1}g \cdot c_{j}(g)\,d_{j}(h)\\
&= \sum_{\ell\in\Lambda}\ell \left[\sum_{g\in\Lambda}\sum_{j = 1}^{n}c_{j}(g)\,d_{j}(\ell^{-1}g)\right]
\end{align}
We now observe that $\overline{S_{2}}\cdot S_{1} = 0$ if and only if the quantity in brackets vanishes $\forall$ $\ell\in\Lambda$.  However, this is precisely the condition (\ref{eq:Commuting_Condition}) required so that $[\mathcal{O}_{1}, \,\ell\cdot\mathcal{O}_{2}] = 0$.  This completes the proof.

A natural consequence of this proposition is the following. Consider $N$ types of operators at each lattice site, and a Hamiltonian of the form:
\begin{align}
H = -\sum_{m}\left[\mathcal{O}^{(1)}_{m} + \mathcal{O}^{(2)}_{m} + \cdots + \mathcal{O}^{(N)}_{m}\right]
\end{align}
where the sum is over all lattice sites. We may now represent each operator $\mathcal{O}^{(i)}$ at a particular lattice site by a stabilizer map $S_{i}$.  Then, all of the operators appearing in the Hamiltonian \emph{mutually commute} (i.e. $[\mathcal{O}^{(i)}_{n}, \mathcal{O}^{(j)}_{m}] = 0$ $\forall$ $i$, $j$, $m$, $n$) if and only if
\begin{align}
\overline{S_{i}}\cdot S_{j} = 0
\end{align}
for any pair $i$, $j$.

From the stabilizer map, we may determine how the action of an arbitrary operator $\mathcal{O}$ on the ground-state creates a pattern of excitations.  Let $S$ be the stabilizer map for an ideal Majorana Hamiltonian with a single operator per lattice site and an $n$-site unit cell. Furthermore, let $P$ be the polynomial representation of some operator $\mathcal{O}$.  We represent $S$ and $P$ as
\begin{align}
S \equiv \sum_{g\in\Lambda}g\cdot\left(\begin{array}{c}
c_{1}(g)\\
c_{2}(g)\\
\vdots\\
c_{n}(g)
\end{array}
\right) \hspace{.12in}
P \equiv \sum_{g\in\Lambda}g\cdot\left(\begin{array}{c}
d_{1}(g)\\
d_{2}(g)\\
\vdots\\
d_{n}(g)
\end{array}
\right)
\end{align}
respectively.  Recall that the condition (\ref{eq:Commuting_Condition}) determines whether the operator $\mathcal{O}$ commutes with a given stabilizer appearing in the Hamiltonian.  Whenever, $\mathcal{O}$ anti-commutes with a stabilizer operator, it creates an excitation at the location of the stabilizer.  The pattern of excitations created by $\mathcal{O}$ is then specified by the expression
\begin{widetext}
\begin{align}
\sum_{h\in\Lambda}h^{-1}\left[\sum_{g\in\Lambda}\sum_{j=1}^{n}c_{j}(g)\,d_{j}(h^{-1}g)\right] = \sum_{h,g\in\Lambda}\sum_{j=1}^{n}h\,g^{-1}c_{j}(g)\,d_{j}(h) = \sum_{j=1}^{n}\left[\sum_{g\in\Lambda}g^{-1}c_{j}(g)\sum_{h\in\Lambda}h\,d_{j}(h)\right] = \bar{S}\cdot P
\end{align}
\end{widetext}
Therefore, we define the \emph{excitation map}
\begin{align}
E \equiv \overline{S}
\end{align}
so that $E\cdot P$ yields a polynomial representation of the pattern of excitations created by the operator $\mathcal{O}$ when acting on the ground-state.

\section{Stabilizer Maps for Ideal Majorana Hamiltonians with a Two-Site Basis} 
\label{app:StabilizerMap}

We now prove the following statement: a topologically-ordered, ideal Majorana Hamiltonian with a two-site basis and a single operator per lattice site, is specified by a stabilizer map is of the form:
\begin{align}
S = \left(\begin{array}{c}
 f(x_{1}, \ldots, x_{d})\\ \\
 \overline{f(x_{1},\ldots,x_{d})}
 \end{array}\right)
\end{align}
with $f(x_{1},\ldots,x_{d})\in R \equiv \mathbb{F}_{2}[x_{1},\ldots,x_{d}]$.  Note that $S$ is uniquely defined only up to monomial multiplication on each of its entries. 
 
We demonstrate this as follows. For an ideal Majorana Hamiltonian with a two-site basis,
let the stabilizer map be of the form $S^{T} = (f, g)$ with $f$, $g\in R$.
%The commutativity condition \eqref{eq:Commutativity} requires that
%\begin{align}
%\bar{f} f + \bar{g} g = 0.
%\end{align}
%A trivial solution to the above equation is given by $\bar{f}f = \bar{g}g = 1$
%so that $f$ and $g$ are monomials.
%Physically, this corresponds to a trivial Hamiltonian describing non-interacting, dimerized Majorana fermions.
%We neglect this solution for the remainder of our analysis.
Recall that the excitation map is given by $E = (\bar{f}, \bar{g})$.
We require on an open lattice that $\ker (E) \cong \mathrm{im}(S)$,
so that any degenerate ground-states of the Hamiltonian are locally indistinguishable.
Note that $\ker(E)$ and $\mathrm{im}(S)$ are defined as 
\begin{align}
&\mathrm{ker}(E) \equiv \Bigg\{\left(\begin{array}{c}\alpha\\ \beta \end{array}\right)\in R^2\,\,\Bigg | \,\,\alpha \bar{f} = \beta \bar{g}\Bigg\}\\
&\mathrm{im}(S) \equiv \Bigg\{\left(\begin{array}{c}c\cdot f\\ c\cdot g \end{array}\right)\,\,\Bigg | \,\,c \in R\Bigg\}
\end{align}

Let $g$ and $f$ take the form $g = h g'$ and $f = h f'$, for some $h \in R$, so that $\mathrm{gcd}(f,g) = h$, and so that $\mathrm{gcd}(g',f') = 1$. In this case, we observe that 
\begin{align}
\left(\begin{array}{c}\overline{g'}\\\\\overline{f'}\end{array}\right) \in \mathrm{ker}(E)
\end{align}
If this element is generated by $\mathrm{im}(S)$, it must be the case that 
\begin{align}
\left(\begin{array}{c}\overline{g'}\\\\\overline{f'}\end{array}\right) = c\left(\begin{array}{c}{f}\\\\{g}\end{array}\right) = c\cdot h\left(\begin{array}{c}{f'}\\\\{g'}\end{array}\right)
\end{align}
for some $c\in R$.  However, this equation is only satisfied if $c \cdot h\cdot\overline{c\cdot h} = 1$, which implies that both $c$ and $h$ must be monomials. As $S$ is only uniquely defined up to monomial multiplication on its entries, we may restrict the remainder of our analysis to stabilizer maps $S = (f, g)^{T}$ with $f$ and $g$ co-prime so that $\mathrm{gcd}(f,g) = 1$.

%An element of $\mathrm{ker}(E)$ takes the general form $\alpha \overline{f'} = \beta \overline{g'}$. As $f'$ and $g'$ are co-prime, a general solution to this equation is $\alpha = r\cdot \overline{g'}$, $\beta = r\cdot \overline{f'}$ for some $r\in R$. This arbitrary element of \mathrm{ker}(E) is generated by $\mathrm{im}(S)$ if 

Since $f$ and $g$ are co-prime, the condition $\alpha\bar{f} = \beta\bar{g}$ for a vector $(\alpha, \beta)^{T}$ to be in $\mathrm{ker}(E)$ is satisfied only if $\alpha = c\cdot \bar{g}$ and $\beta = c\cdot \bar{f}$ for some $c\in R$.  Now consider that in order for $\mathrm{im}(S)\cong\mathrm{ker}(E)$ on an open surface, we must have that
\begin{align}
d\cdot\left(\begin{array}{c} f\\\\g \end{array} \right) = \left(\begin{array}{c}c \cdot \bar{g}\\\\c\cdot \bar{f}\end{array}\right).
\end{align}
for some $d\in R$. This is indeed satisfied if $c = d$, so that $f = \overline{g}$.  Now, we have that the most general form of the stabilizer map (up to monomial multiplication on each entry) is $S = (f, \bar{f})^{T}$ for some $f \in R$ such that $\mathrm{gcd}(f,\overline{f}) = 1$. The commutativity condition $\overline{S}\cdot S = 0$ is trivially satisfied.  This completes the proof.

\end{document}